\documentclass[review]{siamart190516}
\usepackage[utf8]{inputenc}

\usepackage{todonotes}
\usepackage{enumitem}
\usepackage{amsfonts}
\usepackage{amsmath}
\usepackage{algorithm}
\usepackage{algpseudocode}
\usepackage{hyperref}
\usepackage{multirow}
\usepackage{multicol}
\usepackage{subcaption}
\usepackage[capitalize]{cleveref}

\allowdisplaybreaks

\newcommand{\extadd}{\mathbin{\hspace{0.65em}\text{\makebox[0pt]{\resizebox{0.3em}{0.85em}{\(\updownarrow\)}}\raisebox{0.15em}{\makebox[0pt]{\resizebox{1.3em}{0.2em}{\(\leftrightarrow\)}}}}\hspace{0.65em}}}

\newcommand{\ignore}[1]{}
\newcommand{\ylrev}[1]{{\color{black}{#1}}}
\newcommand{\pgrev}[1]{{\color{black}{#1}}}

\usepackage{tikz}
\tikzstyle{cnode} = [draw, circle,scale=.6]
\tikzstyle{level 1} = [level distance=.35\textwidth, sibling distance=1\textwidth]
\tikzstyle{level 2} = [level distance=.35\textwidth, sibling distance=.45\textwidth]
\tikzstyle{level 3} = [level distance=.3\textwidth, sibling distance=.25\textwidth]
\usetikzlibrary{decorations.pathreplacing,patterns}

\title{Sparse Approximate Multifrontal Factorization with Butterfly Compression for High Frequency Wave Equations}

\author{Yang Liu\thanks{Computational Research Division, Lawrence Berkeley National Laboratory, Berkeley, CA, USA.\newline
		(\texttt{\{liuyangzhuan,pghysels,lclaus,xsli\}@lbl.gov})} \and Pieter Ghysels\footnotemark[1] \and Lisa Claus\footnotemark[1] \and Xiaoye Sherry Li\footnotemark[1]}

\nolinenumbers
\begin{document}
	
	\maketitle
	
	\begin{abstract}
		We present a fast and approximate multifrontal solver for large-scale sparse linear systems arising from finite-difference, finite-volume or finite-element discretization of high-frequency wave equations. The proposed solver leverages the butterfly algorithm and its hierarchical matrix extension for compressing and factorizing large frontal matrices via graph-distance guided entry evaluation or randomized matrix-vector multiplication-based schemes. Complexity analysis and numerical experiments demonstrate $\mathcal{O}(N\log^2 N)$ computation and $\mathcal{O}(N)$ memory complexity when applied to an $N\times N$ sparse system arising from 3D high-frequency Helmholtz and Maxwell problems.
	\end{abstract}
	
	\begin{keyword}
		Sparse direct solver, multifrontal method, butterfly algorithm, randomized algorithm, high-frequency wave equations, Maxwell equation, Helmholtz equation, Poisson equation.
	\end{keyword}
	
	\begin{AMS}
		15A23, 65F50, 65R10, 65R20
	\end{AMS}
	
	\section{Introduction}
	Direct solution of large sparse linear systems arsing from e.g., finite-difference, finite-element or finite-volume discretization of partial differential equations (PDE) is crucial for many high-performance scientific and engineering simulation codes. Efficient solution of these sparse systems often requires reordering the matrix to improve numerical stability and fill-in ratios, and performing the computations on smaller but dense submatrices to improve flop performance. Examples include supernodal and multifrontal methods that perform operations on so-called
	supernodes and frontal matrices, respectively~\cite{dufferismanreid2017,davis_2016_survey}.
	For multifrontal methods~\cite{liu92}, the size $n$ of the frontal matrices can grow as
	$n=\mathcal{O}(N^{1/2})$ and $n=\mathcal{O}(N^{2/3})$ for typical 2D and 3D PDEs with $N$ denoting the system size. Performing dense factorization and solution on the frontal matrices requires $\mathcal{O}(n^3)$ operations, yielding the overall complexities of $\mathcal{O}(N^{3/2})$ in 2D and $\mathcal{O}(N^{2})$ in 3D. The same complexities also apply to supernodal methods. 
	
	For many applications arising from wide classes of PDEs, these complexities can be 
	reduced by leveraging algebraic compression tools to exploit rank structures in blocks of the matrix inverse. For example, it can be rigorously shown that, for elliptical PDEs with constant or smooth coefficients, certain off-diagonal blocks in a frontal matrix exhibit low-rankness~\cite{chandrasekaran2010numerical}. Low-rank based fast direct solvers, including $\mathcal{H}$~\cite{hackbusch_sparse_1999,Hackbusch_2003_Hmatrix} and $\mathcal{H}^2$ matrices~\cite{hackbusch_data-sparse_2002}, hierarchically off-diagonal low-rank (HOD-LR) formats~\cite{ambikasaran_mathcalo_2013}, sequentially semi-separable formats \cite{vandebril2005bibliography}, hierarchically semi-separable (HSS) formats \cite{vandebril2005bibliography}, and block low-rank (BLR) formats \ylrev{\cite{Shaeffer2007blr,amestoy2015improving,Theo19performance}}, represent off-diagonal blocks as low-rank products and leverage fast algebras to perform efficient matrix factorization. These methods were first developed \ylrev{to address} dense systems, e.g., arising from boundary element methods, in quasi-linear complexity and have been recently adapted for sparse systems. Examples include solvers coupling $\mathcal{H}$ \cite{zhou2015Hfem}, HOD-LR  \cite{aminfar2016fast}, HSS  
	\cite{xia2013randomized,Wang2016HSS,ghysels2017robust} or BLR  \cite{amestoy2015improving} with multifrontal methods, which we will refer to as rank structured multifrontal methods, and solvers coupling HOD-LR \cite{chadwick2015efficient} or BLR \cite{nies2019testing} with supernodal methods, solvers based on the inverse fast multipole method \cite{Pouransari2017IFMM}, and \ylrev{multifrontal-like solvers based on hierarchical interpolative factorization (HIF) \cite{ho2016hierarchical,li2017distributed}}. Available software packages include STRUMPACK~\cite{ghysels2017robust}, 
	MUMPS~\cite{amestoy2015improving}, and PaStiX~\cite{henon2002pastix,nies2019testing}. 
	It is worth mentioning that many of these methods rely on fast entry evaluation or randomized matrix vector multiplication (matvec) to compress the frontal matrices, without explicitly forming them. Despite differences in the leading constants, implementation details and applicability of these compression formats, in general they lead to quasi-linear complexity direct solvers and preconditioners when applied to many elliptical PDEs. Unfortunately, when applied to wave equations, such as Helmholtz, Maxwell, or Schr\"{o}dinger equations with constant or non-constant coefficients, the frontal matrices exhibit much higher numerical ranks due to the highly oscillatory nature of the numerical Green's function~\cite{Engquist2018Greenfunction} and consequently no asymptotic complexity reduction compared to the exact sparse solvers can be attained. That said, the low-rank based sparse direct solver packages (e.g., STRUMPACK and MUMPS) oftentimes significantly reduce the costs of exact sparse solvers for practical wave equation systems \cite{operto20073d}, \ylrev{and scale quasi-linearly when the 3D computation domain spans less than a few wavelengths.}

	In contrast to low-rank-based algorithms, we consider another algebraic compression tool called butterfly \cite{Eric_1994_butterfly,liu2020butterfly,li_butterfly_2015,Yingzhou_2017_IBF,Pang2020IDBF}, for constructing fast multifrontal methods for wave equations. Butterfly is a multilevel matrix decomposition algorithm well-suited for representing highly oscillatory operators such as Fourier transforms and integral operators \cite{Candes_butterfly_2009,Lexing_SFT_2009,Haizhao_2018_Phase} and special function transforms \cite{Tygert_2010_spherical,bremer2020SHT,Oneil_2010_specialfunction}. When combined with hierarchical matrix techniques, butterfly can also serve as the building block for accelerating iterative methods~\cite{michielssen_multilevel_1996}, direct solvers~\cite{Han_2013_butterflyLU,Han_2017_butterflyLUPEC,Han_2018_butterflyLUDielectric,Liu_2017_HODBF} and preconditioners~\cite{Yang_2020_BFprecondition} for boundary element methods for high-frequency wave equations. These techniques essentially replace low-rank products in the $\mathcal{H}$~\cite{Han_2017_butterflyLUPEC}, $\mathcal{H}^2$~\cite{ying2015directional,Steffen2017directionalh2} and HOD-LR formats~\cite{Liu_2017_HODBF} with butterflies, and leverage fast and randomized butterfly algebra to compute the matrix inverse (for direct solvers and preconditioners). We particularly focus on the butterfly extension of the HOD-LR format~\cite{Liu_2017_HODBF},
	called HOD-BF in this paper. The HOD-BF format yields smaller leading constants and better parallel performance compared to other butterfly-enhanced hierarchical matrix formats. Moreover, HOD-BF can attain an $\mathcal{O}(n\log^2 n)$ compression complexity and an empirical $\mathcal{O}(n^{3/2}\log n)$ inversion complexity given an $n\times n$ HOD-BF compressible dense matrix. HOD-BF has been previously applied to both 2D~\cite{Liu_2017_HODBF} and 3D boundary element methods. 
	
	In this paper, we leverage the HOD-BF format for compressing the frontal matrices in the multifrontal method. \ylrev{The proposed algorithm is formulated as \cref{alg:preconditioner}, which combines several butterfly algorithms at multiple phases of the multifrontal method.} Specifically, any non-root frontal matrix has a $2\times2$ blocked partition, and each block represents numerical Green's function interactions between unknowns residing on planar or crossing planes (or lines). These blocks are compressed as butterfly or HOD-BF by extracting selected matrix entries~\cite{Eric_1994_butterfly,Pang2020IDBF}, from the children frontal matrices and the original sparse matrix. Moreover, the method factorizes the leading diagonal block using the HOD-BF inversion technique \cite{Liu_2017_HODBF} and compute its Schur complement with the randomized butterfly construction algorithm \cite{liu2020butterfly}. \ylrev{It's worth mentioning that efficient integration of butterfly algorithms into the multifrontal method requires several algorithmic innovations. First, butterfly construction of frontal blocks requires sub-sampling matrix entries using proxy rows and columns to maintain quasi-linear complexities. These proxies are selected by combining uniform sampling and nearest neighbor sampling, computed using the graph distance (see \cref{alg:BF_entry}). Second, matrix sub-sampling boils down to extracting entries from the compressed children frontal matrices, implemented as partial matvec in a blocked and parallel fashion (\cref{alg:elem_extract}).}

	Given a frontal matrix of size $n\times n$, the construction and factorization can be performed in $\mathcal{O}(n^{3/2}\log n)$ complexity. Consequently, for a sparse matrix resulting from 2D and 3D high frequency wave equations, the solver can attain an overall complexity of $\mathcal{O}(N)$ and $\mathcal{O}(N\log^2 N)$, respectively. It is worth mentioning the same complexities can be attained for 2D and 3D low-frequency or static PDEs such as the Poisson equation. To the best of our knowledge, the proposed solver represents the first-ever quasi-linear complexity multifrontal solver for high-frequency wave equations \ylrev{in 3D}.

	As a related work, the sweeping preconditioner-based domain decomposition solvers~\cite{taus2019lsweeps} 
	represent another quasi-linear complexity technique for wave equations and impressive numerical results have been reported for both 2D and 3D cases. However, sweeping preconditioners only apply to regular grids and domains and do not work well for domains containing resonant cavities. In comparison, the proposed solver does not suffer from these constraints and applies to wider classes of applications. In addition, the proposed solver can be used inside domain decomposition solvers, which often use multifrontal solvers for their local sparse systems.
	
	The rest of the paper is organized as follows. The multifrontal method is presented in \cref{sec:MF}. The butterfly format, construction and entry extraction algorithms are described in \cref{sec:BF}, followed by their generalization to the HOD-BF format in \cref{sec:HODBF}. The proposed rank structured multifrontal method is detailed in \cref{sec:preconditioner}, including 
	complexity analysis. Numerical results demonstrating the efficiency and applicability of the proposed solver for the 3D Helmholtz, Maxwell, and Poisson equations are presented in \cref{sec:example}, \ylrev{followed by conclusions in \cref{sec:conclude}.}

	\section{Sparse Multifrontal LU Factorization\label{sec:MF}}
	We consider the LU factorization of a sparse matrix $A \in \mathbb{C}^{N \times N}$, as
	$P \left( D_r A D_c Q_c \right) P^\top = LU$,
	where $P$ and $Q_c$ are permutation matrices, $D_r$ and $D_c$ are diagonal row
	and column scaling matrices and $L$ and $U$ are sparse lower and 
	upper-triangular respectively. $Q_c$ aims to maximize the magnitude 
	of the elements on the matrix diagonal. $D_r$ and $D_c$ scale the matrix such that the diagonal 
	entries are one in absolute value and all off-diagonal entries are less than one.
	This step is implemented using the sequential MC64 code~\cite{duff1999design} 
	or the parallel method -- without the diagonal scaling -- described in~\cite{azad2018distributed}.
	The permutation $P$ is applied 
	symmetrically and is used to minimize the fill-in, i.e., the number of non-zero entries in 
	the sparse factors $L$ and $U$. This permutation is computed from the symmetric sparsity
	structure of $A+A^\top$. For large problems the preferred ordering is typically based on 
	the nested dissection heuristic, as implemented in METIS~\cite{karypis1998fast} or
	Scotch.
	
	
	\ignore{ 
		The multifrontal method relies on a structure called
		the elimination tree of $A$, which is a tree with $N$ nodes,
		where the $i$-th node corresponds to the $i$-th column of $L$ and with the parent 
		relations defined by: $\mathrm{parent}(j) = \min\{i: i>j \;\mathrm{and}\; \ell_{ij}\neq
		0\}$, for $j=1,\ldots, N-1$.  This elimination tree serves as a task
		and data-dependency graph for both the factorization and the solution
		process. In practice, nodes are amalgamated: nodes that represent
		columns and rows of the factors with similar structures are grouped
		together in a single node. The resulting tree is called the assembly
		tree.
	} 
	
	The multifrontal method~\cite{duff1983multifrontal} relies on a graph called the assembly tree to guide the computation. Each node $\tau$ of the assembly tree corresponds to a dense frontal matrix $F_\tau$, \ylrev{representing an intermediate dense submatrix in sparse Gaussian elimination, with the following $2 \times 2$ block structure: $F_\tau=\begin{bmatrix}F_{11} & F_{12}\\ F_{21} & F_{22}\end{bmatrix}$. Here, the rows and columns corresponding to the $F_{11}$ block, denoted by index set $I^s_{\tau}$, are called the fully-summed variables, and when the front is constructed, $F_{11}$ is ready for LU factorization. The disjoint index sets $I^s_{\tau}$ form a partition of the index set of $A$ as $\bigcup_\tau I^s_{\tau} = \{1, \dots, N \}$. In the context of nested dissection, the sets $I_{\tau}^s$ correspond to individual vertex separators. The rows and columns corresponding to the $F_{22}$ block, denoted by index set $I^u_\tau$, define the (temporary) Schur complement update blocks that $F_{11}$ contributes to during the multifrontal LU factorization. Let $C_{\tau}=F_{22} - F_{21} F_{11}^{-1} F_{12}$ denote the contribution block, i.e., the Schur complement updated $F_{22}$. If $\nu$ is a child of $\tau$ in the assembly tree, then $I^u_{\nu} \subset \{ I^s_{\tau} \cup I^u_{\tau}\}$; for the root node $t$, $I_t^u \equiv \emptyset$.  Let $\# I^{\text{s}}_\tau$, $\# I^{\text{u}}_\tau$ and $n_\tau = \#I^{\text{s}}_\tau \, + \, \#I^{\text{u}}_\tau$ denote the dimensions of $F_{11}$, $F_{22}$ and $F_\tau$, respectively. Note that $n_\tau$ tends to get bigger toward the root of the assembly tree. When considering a single front, we will omit the $\tau$ subscript.}

	
	The multifrontal method casts the factorization of a sparse matrix
	into a series of partial factorizations \ylrev{and Schur complement updates of the frontal matrices}. It consists in a bottom-up traversal of
	the assembly tree following a topological order. Processing a node
	consists of four steps: 
	\begin{enumerate}[leftmargin=*]
		\setlength{\itemsep}{2pt}%
		\setlength{\parskip}{2pt}%
		\item Assembling the frontal matrix $F_{\tau}$, i.e., combining elements from the sparse matrix $A$ with the
		children's ($\nu_1$ and $\nu_2$) \ylrev{contribution blocks}. 
		This involves a scatter operation and is called extend-add, denoted by $\extadd$:
		\begin{equation} \nonumber
		F_\tau = 
		\begin{bmatrix}
		A(I^s_\tau,I^s_\tau) & A(I^s_\tau,I^u_\tau) \\
		A(I^u_\tau,I^s_\tau)
		\end{bmatrix} \extadd F_{22;\nu_1} \extadd F_{22;\nu_2} 
		= 
		\begin{tikzpicture}[baseline={([yshift=-.8ex]current bounding box.center)}]
		\def\l{1.2}
		\draw (0,0) rectangle (\l/3,\l-\l/3);
		\draw (0,\l) rectangle (\l/3,\l-\l/3);
		\draw (\l/3,\l-\l / 3) rectangle (\l,\l);
		\foreach \x in {0,1,2,...,9} {
			\draw (\x * \l / 30 + \l / 60,\l - \x * \l / 30 - \l / 60) circle (\l / 120);
		}
		\foreach \x in {0,1,...,8} {
			\draw (\x * \l / 30 + \l / 30 + \l / 60,\l - \x * \l / 30 - \l / 60) circle (\l / 120);
		}
		\foreach \x in {0,1,...,8} {
			\draw (\x * \l / 30 + \l / 60,\l - \x * \l / 30 - \l / 60 - \l / 30) circle (\l / 120);
		}
		\draw (\l/2,\l - \l / 8) circle (\l / 120);
		\draw (\l/8,\l - \l / 2) circle (\l / 120);
		\draw (2*\l/3,\l - \l / 6) circle (\l / 120);
		\draw (\l/6,\l - 2*\l / 3) circle (\l / 120);
		\end{tikzpicture}
		\extadd 
		\begin{tikzpicture}[baseline={([yshift=-.8ex]current bounding box.center)}]
		\def\l{.8}
		\draw [fill=lightgray] (0,0) rectangle (\l,\l);
		\end{tikzpicture}
		\extadd
		\begin{tikzpicture}[baseline={([yshift=-.8ex]current bounding box.center)}]
		\def\l{.95}
		\draw [fill=lightgray] (0,0) rectangle (\l,\l);
		\end{tikzpicture}
		\end{equation}
		\item Elimination of the fully-summed variables in the
		$F_{11}$ block, i.e., dense LU factorization with partial pivoting of $F_{11}$.
		\item Updating the off-diagonal blocks $F_{12}$ and $F_{21}$.
		\item \ylrev{Computing the contribution block from the Schur complement update of $F_{22}$:
			$C_{\tau} = F_{22} - F_{21} F_{11}^{-1} F_{12}$. $C_{\tau}$ or $F_{22}$} is temporary storage (pushed on a stack), and can be released as soon as it has been used in the front assembly (step (1)) of the parent node.
	\end{enumerate}
	After the numerical factorization, the lower triangular sparse factor is available in the $F_{21}$ and $F_{11}$ blocks and the upper triangular factor in the $F_{11}$ and $F_{12}$ blocks. These can then be used to very efficiently solve linear systems, using forward and backward substitution. A high-level overview is given in \cref{alg:mf}. 
	
	\ignore{\begin{algorithm}
			\begin{algorithmic}[1]
				\For{nodes $\tau$ in assembly tree in topol. order}
				\State $\footnotesize F_{\tau} \leftarrow 
				\begin{bmatrix}
				A(I_{\tau}^{\text{s}},I_{\tau}^{\text{s}}) & A(I_{\tau}^{\text{s}},I_{\tau}^{\text{u}}) \\
				A(I_{\tau}^{\text{u}},I_{\tau}^{\text{s}}) & 0 \end{bmatrix} \extadd F_{{\nu_1};22} \extadd F_{{\nu_2};22}$ \Comment{assembling: $\nu_1, \nu_2, \ldots$ are $\tau$'s children}
				\State $P_\tau L_\tau U_\tau \leftarrow F_{\tau;11}$ \\
				\hfill \Comment{partial elimination: dense LU w. partial pivoting}
				\State $F_{\tau;12} \leftarrow L_{\tau}^{-1} P_{\tau}^\top F_{\tau;12}$
				\State $F_{\tau;21} \leftarrow F_{\tau;21} U_{\tau}^{-1}$
				\State $F_{\tau;22} \leftarrow F_{\tau;22} - F_{\tau;21} F_{\tau;12}$ \hfill \Comment{Schur update}
				\EndFor
			\end{algorithmic}
			\caption{Sketch of the multifrontal factorization algorithm.}
			\label{alg:mf}
	\end{algorithm}}

	We implemented the multifrontal method in the STRUMPACK
	library~\cite{strumpack_web}, using C++, MPI and OpenMP,
	supporting real/complex arithmetic, single/double precision and
	$32$/$64$-bit integers.
	\ylrev{Note that the pivoting strategy in factorization of $F_{11}$ in STRUMPACK does not use numerical values of $F_{21}$ and $F_{12}$ for ease of implementation.}

	
	For any frontal matrix $F_\tau$ of size $n_\tau$, its LU factorization (only on $F_{11}$) and storage costs scale as $\mathcal{O}(n_\tau^3)$ and $\mathcal{O}(n_\tau^2)$, severely limiting the applicability of the multifrontal method to large-scale PDE problems. In what follows, we leverage the butterfly algorithm and its hierarchical matrix extension for representing frontal matrices and constructing fast sparse direct solvers, particularly for high-frequency wave equations.
	

	\begin{algorithm}
		\textbf{Input:} $A \in \mathbb{R}^{N \times N}$, $b \in \mathbb{R}^{N}$ \\
		\textbf{Output:} $x \approx A^{-1} b$
		\begin{algorithmic}[1]
			\State $A \leftarrow D_r A D_c Q_c$  \hfill \Comment{(optional) col perm \& scaling}
			\State $A \leftarrow P A P^\top$ \hfill \Comment{symm fill-reducing reordering}
			\State Build assembly tree: define $I^{\text{s}}_\tau$ and $I_\tau^{\text{u}}$ for every frontal matrix $F_\tau$
			\For{nodes $\tau$ in assembly tree in topological order}\\
			\Comment{sparse with the children updates extended and added}
			\State $\scriptsize F_{\tau} \leftarrow \begin{bmatrix} A(I_{\tau}^{\text{s}},I_{\tau}^{\text{s}}) & A(I_{\tau}^{\text{s}},I_{\tau}^{\text{u}}) \\
			A(I_{\tau}^{\text{u}},I_{\tau}^{\text{s}}) & 0 \end{bmatrix} \extadd \ylrev{C_{\nu_1}} \extadd \ylrev{C_{\nu_2}}$
			\State $P_\tau L_\tau U_\tau \leftarrow \ylrev{F_{11}}$ \hfill \Comment{LU with partial pivoting}
			\State $\ylrev{F_{12}} \leftarrow L_{\tau}^{-1} P_{\tau}^\top \ylrev{F_{12}}$
			\State $\ylrev{F_{21}} \leftarrow \ylrev{F_{21}} U_{\tau}^{-1}$
			\State $\ylrev{C_{\tau} \leftarrow F_{22} - F_{21} F_{12}}$ \hfill \Comment{Schur update} \label{alg_line:endelseMF}
			\EndFor
			\State $x \leftarrow D_c Q_c P^\top \,\, \text{bwd-solve}\left( \text{fwd-solve} \left(
			P D_r b \right) \right)$
		\end{algorithmic}
		\caption{Sparse multifrontal factorization and solve.}
		\label{alg:mf}
	\end{algorithm}

	\section{Butterfly Algorithms\label{sec:BF}}
	As the building block of our butterfly algorithms, we first present some background regarding the interpolative decomposition (ID).  
	Given a matrix $A\in \mathbb{R}^{m\times n}$, a row ID  represents or approximates $A$ in the low-rank form $UA_{I, :}$, where $U \in \mathbb{R}^{m\times r}$ 
	has bounded entries, $A_{I, :} \in \mathbb{R}^{r\times n}$ contains \ylrev{$r$} rows of $A$, and $r$ is the rank. 
	Symmetrically, a column ID can represent or approximate $A$ column-wise as $A_{:, J} V^\top$ where $V \in \mathbb{R}^{n\times r}$ and $A_{:, J}$ contains $r$ columns of $A$. 
	
	Using an algebraic approach, an ID approximation with a given error threshold can be computed using for instance
	the strong rank-revealing 
	or column-pivoted QR decomposition with typical complexity $\mathcal{O}(rmn)$ (or $\mathcal{O}(r mn\log m )$ in rare cases).%
	\footnote{Note that we use base 2 for the logarithm throughout this paper.}
	In practice, pivoted QR decomposition is more commonly used while entries of the obtained $U$ are mostly bounded (but without theoretical guarantee). Specifically, a \ylrev{row} ID approximation is calculated as follows. Calculate a QR decomposition of $A^\top$ and truncate it 
	with a given error threshold as 
	\begin{equation} \nonumber
	A^\top P = \begin{bmatrix} A_1^\top\ & A_2^\top \end{bmatrix}
	= \begin{bmatrix} Q_1 & Q_2 \end{bmatrix}
	\begin{bmatrix}
	R_{11} & R_{12} \\  & R_{22}
	\end{bmatrix}
	\approx Q_1 
	\begin{bmatrix} R_{11} & R_{12} \end{bmatrix}
	=  A_1^\top \begin{bmatrix} I & R_{11}^{-1}R_{12}\end{bmatrix},
	\end{equation}
	where $P$ is a permutation matrix that indicates the important rows (oftentimes referred to as row skeletons) of $A$. The \ylrev{row} ID approximation is 
	$A \approx P  \begin{bmatrix} 
	I \\
	(R_{11}^{-1}R_{12})^\top
	\end{bmatrix}$ $A_1=UA_1$ where $U$ is the interpolation matrix.

	\subsection{Complementary Low-Rank Property and Butterfly Decomposition}
	We consider the butterfly compression of a matrix $A = K(O,S) \in \mathbb{R}^{m\times n}$ defined by a highly-oscillatory operator $K(\cdot,\cdot)$ and point sets $S$ and $O$. For example, one can think of $K$ as the free-space Green's function for 3D Helmholtz equations, and $S$ and $O$ as sets of Cartesian coordinates representing source and observer points in the Green's function. However, we do not restrict ourselves to analytical functions and geometrical points in this paper. For simplicity, we assume $m=\mathcal{O}(n)$ and we partition $S$ and $O$ using bisection, resulting in the binary trees $\mathcal{T}_S$ and $\mathcal{T}_O$. We number the levels of $\mathcal{T}_O$ and $\mathcal{T}_S$ from the root to the leaves. The root node, denoted by $t$ in $\mathcal{T}_O$ and $s$ in $\mathcal{T}_S$, is at level $0$; its children are at level $1$, etc. All the leaf nodes are at level $L$. At each level $l$, $\mathcal{T}_O$ and $\mathcal{T}_S$ both have $2^l$ nodes. Let $O_\tau$ be the subset of points in $O$ corresponding to node $\tau$ in $\mathcal{T}_O$. Furthermore, for any non-leaf node $\tau\in\mathcal{T}_O$ with children $\tau_1$ and $\tau_2$, $O_{\tau_1} \cup O_{\tau_2} = O_\tau$ and $O_{\tau_1} \cap O_{\tau_2} = \emptyset$. With a slight abuse of notation, we also use $\tau_i,~ i=1,\ldots,2^l$ to denote all nodes at level $l$ of $\mathcal{T}_O$. The same properties hold true for the partitioning of $S$. 
	
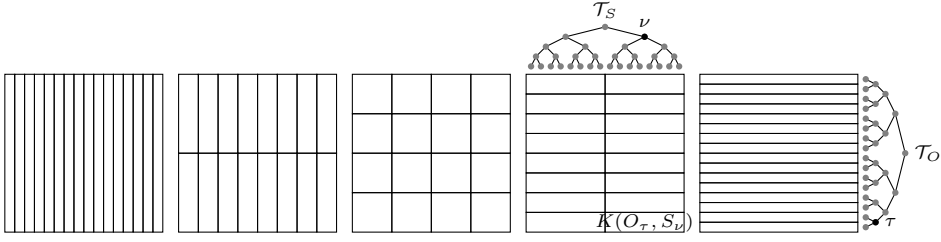
\begin{figure}
  \begin{center}
      \begin{tikzpicture}[scale=2.1]
        \def\myshift{.1}
        \def\s{1/16}
        \def\o{1/8}
        \def\q{1/4}
        \def\h{1/2}
        \foreach \x in {0,1,...,15} {
          \draw (\x * \s,0) rectangle ++(\s,1);
        }
        \tikzset{shift={(1+\myshift,0)}}
        \foreach \x in {0,1,...,7} {
          \foreach \y in {0,1} {
            \draw (\x * \o,\y * \h) rectangle ++(\o,\h);
          }
        }
        \tikzset{shift={(1+\myshift,0)}}
        \foreach \x in {0,1,2,3} {
          \foreach \y in {0,1,2,3} {
            \draw (\x * \q,\y * \q) rectangle ++(\q,\q);
          }
        }
        \tikzset{shift={(1+\myshift,0)}}
        \foreach \x in {0,1} {
          \foreach \y in {0,1,...,7} {
            \draw (\x * \h,\y * \o) rectangle ++(\h,\o);
          }
        }
        \node[auto,font=\scriptsize] at (\h+\q,\s){$K(O_{\tau},S_{\nu})$};
        \tikzset{shift={(1+\myshift,0)}}
        \foreach \x in {0} {
          \foreach \y in {0,1,...,15} {
            \draw (0,\y * \s) rectangle ++(1,\s);
          }
        }
        
        \tikzset{shift={(1+\myshift/2,0)}}
        \foreach \x in {0,2,...,15} {
          \draw (0,\x * \s + \s /2) -- (\s,\x * \s + \s);
          \draw (0,\x * \s + 3*\s / 2) -- (\s,\x * \s + \s);
        }
        \foreach \x in {0,4,...,15} {
          \draw (\s,\x * \s + \s) -- (2*\s,\x * \s + 2*\s);
          \draw (\s,\x * \s + 3*\s) -- (2*\s,\x * \s + 2*\s);
        }
        \foreach \x in {0,8,...,15} {
          \draw (2*\s,\x * \s + 2*\s) -- (3*\s,\x * \s + \q);
          \draw (2*\s,\x * \s + \q+\s+\s) -- (3*\s,\x * \s + \q);
        }
        \draw (3*\s, \q) -- (4*\s,\h);
        \draw (3*\s, 3*\q) -- (4*\s,\h);
        \foreach \x in {0,1,...,15} {
          \fill [gray] (0,\x * \s + \s /2) circle (.02);
        }
        \foreach \x in {0,2,...,15} {
          \fill [gray] (\s,\x * \s + \s) circle (.02);
        }
        \fill [black] (\s,\s) circle (.02);
        \node[right,font=\scriptsize] at (\s,\s){$\tau$};
        \foreach \x in {0,4,...,15} {
          \fill [gray] (2*\s,\x * \s + 2*\s) circle (.02);
        }
        \foreach \x in {0,8,...,15} {
          \fill [gray] (3*\s,\x * \s + \q) circle (.02);
        }
        \fill [gray] (4*\s,\h) circle (.02);
        \node[right,font=\scriptsize] at (4*\s,\h){$\mathcal{T}_O$};

        \tikzset{shift={(-2-1.5*\myshift,1+\myshift/2)}}
        \foreach \x in {0,2,...,15} {
          \draw (\x * \s + \s /2,0) -- (\x * \s + \s,\s);
          \draw (\x * \s + 3*\s / 2,0) -- (\x * \s + \s,\s);
        }
        \foreach \x in {0,4,...,15} {
          \draw (\x * \s + \s,\s) -- (\x * \s + 2*\s,2*\s);
          \draw (\x * \s + 3*\s,\s) -- (\x * \s + 2*\s,2*\s);
        }
        \foreach \x in {0,8,...,15} {
          \draw (\x * \s + 2*\s,2*\s) -- (\x * \s + \q,3*\s);
          \draw (\x * \s + \q+\s+\s,2*\s) -- (\x * \s + \q,3*\s);
        }
        \draw (\q,3*\s) -- (\h,4*\s);
        \draw (3*\q,3*\s) -- (\h,4*\s);
        \foreach \x in {0,1,...,15} {
          \fill [gray] (\x * \s + \s /2,0) circle (.02);
        }
        \foreach \x in {0,2,...,15} {
          \fill [gray] (\x * \s + \s,\s) circle (.02);
        }
        \foreach \x in {0,4,...,15} {
          \fill [gray] (\x * \s + 2*\s,2*\s) circle (.02);
        }
        \foreach \x in {0,8,...,15} {
          \fill [gray] (\x * \s + \q,3*\s) circle (.02);
        }
        \fill [black] (8*\s+\q,3*\s) circle (.02);
        \node[above,font=\scriptsize] at (8*\s+\q,3*\s){$\nu$};
        \fill [gray] (\h,4*\s) circle (.02);
        \node[above,font=\scriptsize] at (\h,4*\s){$\mathcal{T}_S$};
      \end{tikzpicture}
  \end{center}
 \caption{For a $4$-level butterfly decomposition, the complementary low-rank property states that each of the illustrated sub-blocks $K(O_{\tau},S_{\nu}), \tau \in \mathcal{T}_O, \nu \in \mathcal{T}_S$ are low-rank.
 \label{fig:complementary_rank}}
\vspace{-30pt}
\end{figure}

	$A = K(O, S)$ satisfies the \textit{complementary low-rank property} if for any level $0\leq l\leq L$, node $\tau$ at level $l$ of $\mathcal{T}_O$ and a node $\nu$ at level $(L-l)$ of $\mathcal{T}_S$, the subblock $K(O_\tau, S_\nu)$ is numerically low-rank with rank $r_{\tau,\nu}$ bounded by a small number $r$; $r$ is called the (maximum) butterfly rank. For simplicity, we assume constant butterfly ranks $r=\mathcal{O}(1)$ throughout \cref{sec:BF,sec:HODBF}. As explained in Section 4.5 of \cite{liu2020butterfly}, low complexities for butterfly construction, multiplication, inversion and storage can still be achieved even for certain cases of non-constant ranks, e.g., $r=\mathcal{O}(\log n)$ or $r=\mathcal{O}(n^{1/4})$. We will further discuss the non-constant rank case in \cref{ssec:cc}. The complementary low-rank property is illustrated in \cref{fig:complementary_rank}. At any level $l$ of $\mathcal{T}_O$, $K(O_\tau, S_\nu)$ with all nodes $\tau$ at level $l$ of $\mathcal{T}_O$ and nodes $\nu$ at level $(L-l)$ of $\mathcal{T}_S$  (referred to as the blocks at level $l$) form a non-overlapping partitioning of $K(O, S)$. \ylrev{Note that in \cref{fig:complementary_rank}, the rows and columns may have been reordered such that $O_\tau$ and $S_\nu$ correspond to contiguous indices.} 
	
\begin{figure}
  \begin{center}
    \begin{tikzpicture}[scale=3.1]
      \def\myshift{.53}
      \def\r{1/32}
      \def\rr{1/16}
      \def\h{1/2}
      \def\q{1/4}
      \def\o{1/8}
      \foreach \x in {0,1,...,15} {
        \fill [gray] (\x*\r,1-\x*\rr) rectangle ++(\r,-\rr);
      };
      \draw (0,0) rectangle (\h,1.0);
      \node[above] at (\q,1){$U^4$};
      \tikzset{shift={(\myshift,0)}}
      \foreach \x in {0,1,...,7} {
        \fill [gray] (\x*\r,1-\x*\rr) rectangle ++(\r,-\rr);
        \fill [gray] (\q+\x*\r,1-\x*\rr) rectangle ++(\r,-\rr);
      };
      \draw (0,1) rectangle (\h,\h);
      \node[above] at (\q,1){$R^3$};
      \tikzset{shift={(\myshift,0)}}
      \foreach \x in {0,1,...,3} {
        \fill [gray] (\x*\r,1-\x*\rr) rectangle ++(\r,-\rr);
        \fill [gray] (\o+\x*\r,1-\x*\rr) rectangle ++(\r,-\rr);
        \fill [gray] (\q+\x*\r,1-\q-\x*\rr) rectangle ++(\r,-\rr);
        \fill [gray] (\h-\o+\x*\r,1-\q-\x*\rr) rectangle ++(\r,-\rr);
      };
      \draw (0,1) rectangle (\h,\h);
      \node[above] at (\q,1){$R^2$};
      \draw (0,1) rectangle (\q,1-\q);
      \draw (\h,\h) rectangle (\q,1-\q);
      \tikzset{shift={(\myshift,0)}}
      \draw (0,1) rectangle (\h,\h);
      \node[above] at (\q,1){$B^2$};
      \foreach \x in {0,1,2,3} {
        \foreach \y in {0,1,2,3} {
          \fill [gray] (\x*\o+\y*\r,1-\y*\o-\x*\r) rectangle ++(\r,-\r);
          \draw (\x*\o,1-\y*\o) rectangle ++(\o,-\o);
        }
      }
      \tikzset{shift={(\myshift,0)}}
      \foreach \x in {0,1,...,3} {
        \fill [gray] (\x*\rr,1-\x*\r) rectangle ++(\rr,-\r);
        \fill [gray] (\x*\rr,1-\o-\x*\r) rectangle ++(\rr,-\r);
        \fill [gray] (\q+\x*\rr,1-\q-\x*\r) rectangle ++(\rr,-\r);
        \fill [gray] (\q+\x*\rr,1-\h+\o-\x*\r) rectangle ++(\rr,-\r);
      };
      \node[above] at (\q,1){$W^2$};
      \draw (0,1) rectangle (\h,\h);
      \draw (0,1) rectangle (\q,1-\q);
      \draw (\h,\h) rectangle (\q,1-\q);
      \tikzset{shift={(\myshift,0)}}
      \foreach \x in {0,1,...,7} {
        \fill [gray] (\x*\rr,1-\x*\r) rectangle ++(\rr,-\r);
        \fill [gray] (\x*\rr,1-\q-\x*\r) rectangle ++(\rr,-\r);
      };
      \draw (0,1) rectangle (\h,\h);
      \node[above] at (\q,1){$W^1$};
      \tikzset{shift={(\myshift,0)}}
      \foreach \x in {0,1,...,15} {
        \fill [gray] (\x*\rr,1-\x*\r) rectangle ++(\rr,-\r);
      };
      \draw (0,1.0) rectangle (1,\h);
      \node[above] at (\h,1){$V^0$};
    \end{tikzpicture}
  \end{center}
 \caption{Illustration of a $4$-level butterfly representation. For a butterfly representation, we typically put the inner factor $B^l$ at the center level ($l = l_c = L/2$). \label{fig:BF_hybrid}}
 \vspace{-20pt}
\end{figure}
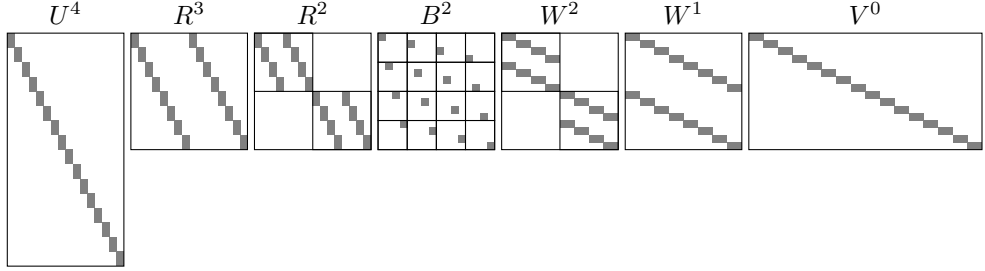

	For any level $l$, we can compress $K(O_\tau, S_\nu)$ via row-wise and column-wise ID as
	\begin{equation}
	K(O_\tau, S_\nu) \approx U_{\tau,\nu} K(\bar{O}_\tau, \bar{S}_\nu) V_{\tau,\nu}^\top = U_{\tau,\nu} B_{\tau,\nu} V_{\tau,\nu}^\top.\label{eqn:IDbf}
	\end{equation}
	Here, $\bar{O}_\tau$ represents skeleton rows (constructed from $O_\tau$), $\bar{S}_\nu$ represents skeleton columns (constructed from $S_\nu$), and $B_{\tau,\nu}$ is the {\it skeleton matrix}. The row and column interpolation matrices $U_{\tau,\nu}$ and $V_{\tau,\nu}$ are defined as 
	\begin{equation}\label{eqn:nested_basis}
	U_{\tau,\nu} =
	\begin{bmatrix}
	U_{\tau_1, p_\nu} & \\
	& U_{\tau_2, p_\nu}
	\end{bmatrix}
	R_{\tau,\nu}, \qquad 
	V_{\tau,\nu}^\top =
	W_{\tau,\nu}\begin{bmatrix}
	V_{p_\tau, \nu_1}^\top & \\
	& V_{p_\tau, \nu_2}^\top
	\end{bmatrix}.
	\end{equation}
	where $R_{\tau,\nu}$ and $W_{\tau,\nu}$ are referred to as the {\it transfer matrices}, and $p_\tau, p_\nu$ denote the parent nodes of $\tau,\nu$. Oftentimes we choose a center level $l=l_c=L/2$ \ylrev{for explicitly using the skeleton matrices $B_{\tau,\nu}$ in \cref{eqn:IDbf}}, and the butterfly representation of $K(O, S)$, \ylrev{referred to the hybrid butterfly representation in \cite{liu2020butterfly}}, is constructed as, 
	\begin{equation}
	K(O, S) =\big(U^LR^{L-1}R^{L-2}\ldots R^{l_c}\big)B^{l_c}\big(W^{l_c}W^{l_c-1}\ldots W^1V^0\big)\label{eqn:hybrid_butterfly_mat}
	\end{equation}
	where $U^L=\mathrm{diag}(U_{\tau_1,s},\ldots,U_{\tau_{2^L},s})$ consists of column basis matrices at level $L$, and each factor $R^{l},l=L-1,\ldots,l_c$ is block diagonal consisting of diagonal blocks $R_{\nu}$ for all nodes $\nu$ at level $L-l-1$ of $\mathcal{T}_S$
	\begin{equation}
	R_\nu=
	\begin{bmatrix}
	\text{diag}(R_{\tau_1, \nu_1}, \dots, R_{\tau_{2^{l}},\nu_1}) & 
	\text{diag}(R_{\tau_1, \nu_2}, \dots, R_{\tau_{2^{l}},\nu_2})
	\end{bmatrix}  .
	\end{equation}
	Here, $\tau_1, \tau_2, \ldots, \tau_{2^{l}}$ are the nodes at level $l$ of $\mathcal{T}_O$ and $\nu_1$, $\nu_2$ are children of $\nu$. Similarly, $V^0=\mathrm{diag}(V^\top_{t,\nu_1},\ldots,V^\top_{t,\nu_{2^L}})$ with $t$ denoting the root of $\mathcal{T}_O$, and the block-diagonal inner factors $W^{l}, l=1,\ldots,l_c$ have blocks $W_{\tau}$ for all nodes $\tau$ at level $l-1$ of $\mathcal{T}_T$
	\begin{align}
	W_\tau=
	\begin{bmatrix}
	\text{diag}(W_{\tau_1,\nu_1}, \dots, W_{\tau_1,\nu_{2^{L-l}}}) \\
	\text{diag}(W_{\tau_2,\nu_1}, \dots, W_{\tau_2,\nu_{2^{L-l}}})
	\end{bmatrix}
	\end{align}
	Here, $\nu_1, \nu_2, \ldots, \nu_{2^{L-l}}$ are the nodes at level $L-l$ of $\mathcal{T}_S$ and $\tau_1$, $\tau_2$ are children of $\tau$. Moreover, the inner factor $B^{l_c}$ consists of blocks $B_{\tau,\nu}$ at level $l_c$ in \cref{eqn:IDbf}. For simplicity assuming $r_{\tau,\nu}=r$, $B^{l_c}$ is a $p\times q$ block-partitioned matrix with each block of size $qr\times pr$; the $(i,j)$ block is a $q\times p$ block-partitioned matrix with each block of size $r\times r$, among which the only nonzero block is the $(j,i)$ block and equals $B_{\tau_i,\nu_j}$. We call \cref{eqn:hybrid_butterfly_mat} a butterfly representation of $A=K(O,S)$, or simply, a butterfly. These structures are illustrated in \cref{fig:BF_hybrid}. Once factorized in the form of \cref{eqn:hybrid_butterfly_mat}, the storage and application costs of a matrix-vector product scale as $\mathcal{O}(n\log n)$. \ylrev{Na\"ive} butterfly construction of \cref{eqn:hybrid_butterfly_mat} requires $\mathcal{O}(n^2)$ operations. However, we consider two scenarios that allow fast butterfly construction: when individual elements of $A$ can be quickly computed, \cref{ssec::extraction-bf-construction}, or when $A$ can be applied efficiently to a set of random vectors, \cref{ssec::random-bf-construction}.
	
	\subsection{Butterfly Construction using Matrix Entry Evaluation\label{ssec::extraction-bf-construction}}
	Oftentimes fast access to any entry of $A$ is available, e.g.,
	when the matrix entry has a closed-form expression, or $A$ has been
	stored in full or compressed forms. If any entry of $A$ can be computed in less than e.g., $\mathcal{O}(\log n)$ operations, the butterfly construction cost can be reduced to quasi-linear time.
	
	Starting from level $L$ of $\mathcal{T}_O$, we need to compute the interpolation matrices $U_{\tau_i,s}$ via row ID such that $K(O_{\tau_i}, S_{s}) = U_{\tau_i,s} K(\bar{O}_{\tau_i}, S_{s})$, $i=1,\ldots,2^{L}$ for the root node $s$ of $\mathcal{T}_S$. Note that it is expensive to perform such direct computation as there are $2^L=\mathcal{O}(n)$ IDs each requiring at least $\mathcal{O}(m)$ operations. Instead, we consider using proxy columns to reduce the ID costs. Specifically, we choose $\mathcal{O}(r)$ columns ${S}_{\tau_i,s}$ from $S_s$ and compute $U_{\tau_i,s}$ from $K(O_{\tau_i}, S_{\tau_i,s}) = U_{\tau_i,s} K(\bar{O}_{\tau_i}, S_{\tau_i,s})$. \ylrev{Recall that for any submatrix $K(O_{\tau},S_{\nu})$, we use $S_{\tau,\nu}$ and $\bar{O}_{\tau}$ as the proxy columns and skeleton rows, respectively. Similarly, we use $O_{\tau,\nu}$ and $\bar{S}_{\tau}$ as the proxy rows and skeleton columns, respectively.} There exists several options on how to choose the proxy columns, including uniform, random or Chebyshev samples~\cite{Yingzhou_2017_IBF}. However, uniform or random samples often yield inaccuracies when the operator represents interactions between close-by spatial domains, and Chebyshev samples only apply to regular spatial domains. Instead we pick $(\alpha+k_{\rm nn})|O_{\tau_i}|$ columns \ylrev{$S_{\tau_i,s}$} with $\alpha|O_{\tau_i}|$ uniform samples ($\alpha$ is an oversampling factor) and $k_{\rm nn}$ nearest points per row using a certain distance metric, see also \cref{ssec:neighbor_sampling} for its application to frontal matrix compression. 
	
	At any level $l=L-1,\ldots,l_c$, we can compute the transfer matrix $R_{\tau, \nu}$ for node $\tau$ at level $l$ of $\mathcal{T}_O$ and node $\nu$ at level $L-l$ of $\mathcal{T}_S$, from  
	\begin{equation}\label{eqn:transfer_matrix}
	K(O_\tau, S_\nu) 
	= 
	\begin{bmatrix}
	U_{\tau_1, p_\nu} &  \\
	& U_{\tau_2, p_\nu}
	\end{bmatrix}
	\begin{bmatrix}
	K(\bar{O}_{\tau_1}, S_\nu) \\
	K(\bar{O}_{\tau_2}, S_\nu)
	\end{bmatrix}= 
	\begin{bmatrix}
	U_{\tau_1, p_\nu} &  \\
	& U_{\tau_2, p_\nu}
	\end{bmatrix}
	R_{\tau, \nu}K(\bar{O}_{\tau}, S_\nu). 
	\end{equation}
	From \cref{eqn:transfer_matrix}, the transfer matrix $R_{\tau, \nu}$ can be computed as the interpolation matrix in the row ID of $K(\bar{O}_{\tau_1}\cup\bar{O}_{\tau_2}, S_\nu)$. Just like level $L$, we choose $(\alpha+k_{\rm nn})|\bar{O}_{\tau_1}\cup\bar{O}_{\tau_2}|$ columns $S_{\tau,\nu}$ from $S_\nu$ as the proxy columns to compute $R_{\tau, \nu}$.  
	
	Similarly, we compute the interpolation matrices $V_{\tau, \nu}$
	at level $0$ and transfer matrices $W_{\tau, \nu}$ at levels $l=1,\ldots,l_c$ using column IDs with uniform and nearest neighboring sampling. Finally, the skeleton matrices $B_{\tau,\nu}$ are directly assembled at center level $l_c$.  
	
	The above-described process is summarized as BF\_entry\_eval($A$) (\cref{alg:BF_entry}). The algorithm computes the butterfly structure of Figure~\ref{fig:BF_hybrid} in an outer-to-inner sequence. \ylrev{The left column represents construction from the leftmost level to the middle level, the right column represents construction from rightmost level to the middle level.} Note that at each level $l=0,\ldots,L$ one needs to extract $\mathcal{O}(n)$ submatrices of size $\mathcal{O}(r)\times \mathcal{O}(r)$ using the element extraction function extract($\mathcal{L},A$) at lines \ref{line:extractL}, \ref{line:extractR}, \ref{line:extractC}. \ylrev{Note that this function is called only $L+2$ times to improve the computational efficiency of BF\_entry\_eval($A$).} This function can efficiently compute a list of submatrices indexed by a list of (rows, columns) index sets $\mathcal{L} = \left\{(X_1,Y_1),(X_2, Y_2),\dots \right\}$, \ylrev{where index sets $X_i$ and $Y_i$ respectively correspond to the row and column indices of $i$th submatrix, which are identified by the proxy rows/columns and skeleton columns/rows at the previous level $l$. Note that for each $(X,Y)\in \mathcal{L}$, $X$ corresponds to one $\tau$ at level $l$ of $\mathcal{T}_O$, and $Y$ corresponds to one $\nu$ at level $L-l$ of $\mathcal{T}_S$.} When $A$ has a closed-form expression or has been stored in full, extract($\mathcal{L},A$) takes $\mathcal{O}(n)$ time and the butterfly construction requires $\mathcal{O}(n\log n)$ time; when $A$ has been computed in some compressed form (e.g., as summation of two butterflies), extract($\mathcal{L},A$) often takes $\mathcal{O}(n\log n)$ time and the butterfly construction requires $\mathcal{O}(n\log^2 n)$ time. As we will see, the latter case appears when compressing the frontal matrices and we describe the extract function with compressed $A$ in \cref{ssec::elem-extraction}.  
	
	\begin{algorithm}
		\caption{BF\_entry\_eval($A$): Butterfly construction of matrix $A$ with entry evaluation.}
		\label{alg:BF_entry}
		\hspace*{\algorithmicindent} \textbf{Input:} A routine extract($\mathcal{L}, A$) to extract a list of sub-matrices of $A$ with $\mathcal{L}$ denoting the list of (rows, columns) index sets, an over-sampling parameter $\alpha$, nearest neighbor parameter $k_{\rm nn}$, ID with a tolerance $\varepsilon$ named ID$_\varepsilon$, and binary partitioning trees $\mathcal{T}_S$ and $\mathcal{T}_O$ of $L$ levels. \ylrev{$O_{\tau,\nu}$ and $S_{\tau,\nu}$ are proxy rows and columns from nearest neighbor and uniform sampling, $\bar{O_{\tau}}$ and $\bar{S}_{\nu}$ are skeleton rows and columns from ID. }\\
		\hspace*{\algorithmicindent} \textbf{Output:} $A=K(O,S)\approx (U^LR^{L-1}R^{L-2}\ldots R^{l_c})B^{l_c}(W^{l_c}W^{l_c-1}\ldots W^1V^0)$ with $l_c=L/2$
		\vspace{-10pt}
		\begin{multicols}{2}
			\begin{algorithmic}[1]	
				\For{$l=L$ to $l_c$}\Comment{\ylrev{Left to middle}}
				\State $\mathcal{L} \leftarrow \{ \}$
				\For{$(\tau,\nu)$ at ($l, L\!-\!l$) of $(\mathcal{T}_O, \mathcal{T}_S)$}
				\If{$l=L$}
				\State $\mathcal{L} \leftarrow \left\{ \mathcal{L}, (O_{\tau},S_{\tau,\nu})\right\}$ with \\
				\hfill $|S_{\tau,\nu}|=(\alpha+k_{\rm nn})|O_{\tau}|$
				\Else
				\State \!\!\!\!\!\!\!\!\!\!\!\!\!$\mathcal{L} \leftarrow \left\{ \mathcal{L}, (\bar{O}_{\tau_1}\cup\bar{O}_{\tau_2}, S_{\tau,\nu})\right\} \mathrm{with}$\\
				\hfill $|S_{\tau,\nu}|=(\alpha+k_{\rm nn})|\bar{O}_{\tau_1}\cup\bar{O}_{\tau_2}|$
				\EndIf
				\EndFor
				\State \ylrev{$\{\forall (X\!,\!Y\!)\!\in\! \mathcal{L} \!:\! K(X\!,\!Y\!)\!\}$ \\
					~~~~~~~~~~$\leftarrow {\rm extract}(\mathcal{L}, A)$} \label{line:extractL}
				\For{$(X, Y) \!\in\! \mathcal{L}~(\mathrm{corresp.}~(\tau, \nu))$}
				\State \!\!\!\!\!\!\!\!\!\!\!\!\!\!\!$U_{\tau,\nu} ({\rm  or~} R_{\tau, \nu}), \bar{O}_{\tau} \leftarrow \mathrm{ID}_\varepsilon ~\mathrm{of}~ K(X,Y)$
				\EndFor	
				\EndFor
				\columnbreak
				
				\For{$l=0$ to $l_c$}\Comment{\ylrev{Right to middle}}
				\State $\mathcal{L} \leftarrow \{ \}$
				\For{$(\tau, \nu)$ at $(l,L\!-\!l)$ of $(\mathcal{T}_O, \mathcal{T}_S)$}
				\If{$l=0$}
				\State $\mathcal{L} \leftarrow \left\{ \mathcal{L}, (O_{\tau,\nu}, S_{\nu}) \right\}$ with \\ \hfill $|O_{\tau,\nu}|=(\alpha+k_{\rm nn})|S_{\nu}|$
				\Else
				\State \!\!\!\!\!\!\!\!\!\!\!\!\!$\mathcal{L} \leftarrow \left\{ \mathcal{L}, (O_{\tau,\nu}, \bar{S}_{\nu_1}\cup\bar{S}_{\nu_2})\right\} \mathrm{with}$\\
				\hfill $|O_{\tau,\nu}|=(\alpha+k_{\rm nn})|\bar{S}_{\nu_1}\cup\bar{S}_{\nu_2}|$
				\EndIf
				\EndFor
				\State \ylrev{$\{\forall (X\!,\!Y\!)\!\in\! \mathcal{L} \!:\! K(X\!,\!Y\!)\!\}$ \\
					~~~~~~~~~~$\leftarrow {\rm extract}(\mathcal{L}, A)$} \label{line:extractR}
				\For{$(X, Y) \!\in\! \mathcal{L}~(\mathrm{corresp.}~(\tau, \nu))$}
				\State \!\!\!\!\!\!\!\!\!\!\!\!\!\!\!$V_{\tau,\nu} ({\rm  or~} W_{\tau, \nu}), \bar{S}_{\nu} \leftarrow ~\mathrm{ID}_\varepsilon ~\mathrm{of}~ K(X,Y)$
				\EndFor	
				\EndFor		
				\State $\mathcal{L} \leftarrow \left\{ \forall \, \tau,\nu {\rm \, at \, level \,} l_c : (\bar{O}_{\tau}, \bar{S}_{\nu}) \right\}$ 
				\State $\left\{ \forall \, \tau\!,\!\nu {\rm \, at \, level \,} l_c \!:\! B_{\tau,\nu} \right\} \!\leftarrow\! {\rm extract}(\mathcal{L}, A)$ \label{line:extractC}
			\end{algorithmic}
		\end{multicols}	
		\vspace{-20pt}
	\end{algorithm}

	\subsection{Randomized Matrix-Free Butterfly Construction\label{ssec::random-bf-construction}}
	
	When fast matrix entry evaluation for a matrix $A$ is not available, but the matrix can be applied to arbitrary vectors in quasi-linear time, typically $\mathcal{O}(n \log n)$, the randomized matrix-free butterfly methods from~\cite{guo2017butterfly} and~\cite{liu2020butterfly} can be used. We use the method from~\cite{liu2020butterfly}, which, given a $\mathcal{O}(n \log {n})$ matrix-vector product, requires $\mathcal{O}(n^{3/2} \log n)$ operations and $\mathcal{O}(n \log n)$ storage. We refer the reader to~\cite{liu2020butterfly} for the details of this algorithm. Throughout this paper, we name this algorithm as BF\_random\_matvec($A$).

	\begin{algorithm}
		\caption{${\rm extract\_BF}(\mathcal{L},A)$: Extraction of a list $\mathcal{L}$ of sub-matrices of a butterfly-compressed matrix $A$.}
		\label{alg:elem_extract}
		\hspace*{\algorithmicindent} \textbf{Input:} $A=(U^LR^{L-1}R^{L-2}\ldots R^{l_c})B^{l_c}(W^{l_c}W^{l_c-1}\ldots W^1V^0)\approx K(O,S)$. A list of (rows, columns) index sets $\mathcal{L} = \{(X_1, Y_1), \dots \}$. \\
		\hspace*{\algorithmicindent} \textbf{Output:} $\forall (X,Y)\in \mathcal{L}: K(X,Y)$. 	
		\vspace{-10pt}
		\begin{multicols}{2}	
			\begin{algorithmic}[1]
				\For{$(X, Y)$ in $\mathcal{L}$}
				\For{$l=0$ to $L$}
				\State Generate a list $\mathcal{L}_l$ of $(\tau,\nu)$ at level $(l,L-l)$ of $(\mathcal{T}_O, \mathcal{T}_S)$ with $X\cap O_\tau \not= \emptyset$ and $Y\cap S_\nu \not= \emptyset$\label{line:listl}
				\EndFor  
				
				\For{$l=L$ to $l_c$}
				\For{($\tau$,$\nu$) in $\mathcal{L}_l$}\label{line:loopl}
				\If{$l=L$}
				\State \!\!\!\!\!\!\!\!\!\!\!\!\!\!$E^l_{\tau,\nu}=U_{\tau,\nu}(I,:)$ 
				\State \!\!\!\!\!\!\!\!\!\!\!\!\!\!\!\!\!\!\!\!\!\!\!\!\!\!\!\!$I \mathrm{~corresponds ~to ~points ~in~} X\!\cap O_\tau$ 
				\Else
				\State \!\!\!\!\!\!\!\!\!\!\!\!\!\!$E^l_{\tau,\nu}=[E^{l+1}_{\tau_1,p_\nu},E^{l+1}_{\tau_2,p_\nu}]R_{\tau,\nu}$ \label{line:ER}
				\EndIf
				\EndFor
				\EndFor		
				\columnbreak
				
				\For{$l=0$ to $l_c$}
				\For{($\tau$,$\nu$) in $\mathcal{L}_l$}\label{line:loopr}
				\If{$l=0$}
				\State \!\!\!\!\!\!\!\!\!\!\!\!\!\!$F^l_{\tau,\nu}=V_{\tau,\nu}^\top(:,J)$ 
				\State \!\!\!\!\!\!\!\!\!\!\!\!\!\!\!\!\!\!\!\!\!\!\!\!\!\!\!\!$J ~\mathrm{corresponds ~to ~points ~in} ~Y\cap S_\nu$
				\Else                                                      
				\State \!\!\!\!\!\!\!\!\!\!\!\!\!\!$F^l_{\tau,\nu}=W_{\tau,\nu}[F^{l-1}_{p_\tau,\nu_1};F^{l-1}_{p_\tau,\nu_2}]$\label{line:WF}
				\EndIf
				\EndFor                                                                                                     
				\EndFor		
				\State \!\!\!\!\!\!\!\!\!\! $K(X,Y)\!\leftarrow\! E^{l_c}_{\tau,\nu}B^{l_c}_{\tau,\nu}F^{l_c}_{\tau,\nu} \, \forall (\tau,\nu) \in \mathcal{L}_{l_c}$\label{line:EBF}
				\EndFor	
				
			\end{algorithmic}
		\end{multicols}
		\vspace{-10pt}    
	\end{algorithm}

	\subsection{Extracting Elements from a Butterfly Matrix\label{ssec::elem-extraction}}
	As explained in more detail in \cref{sec:preconditioner}, incorporating butterfly compression in the sparse solver requires both the BF\_entry\_eval and BF\_random\_matvec algorithms. 
	In one step of the multifrontal algorithm, a subblock of a frontal matrix will be constructed as a butterfly matrix using the BF\_entry\_eval \cref{alg:BF_entry}. Since fronts are constructed as a combination (extend-add) of other smaller fronts, the extract routine used in BF\_entry\_eval will need to extract a list of submatrices from other fronts which might already be compressed using butterfly. Therefore it is critical for performance to have an efficient algorithm to extract a list of submatrices from a butterfly matrix. This is presented as extract\_BF in \cref{alg:elem_extract}.
	
	
	Given an $m\times n$ butterfly matrix $A\approx K(O,S)$ and a list of (rows, columns) index sets $\mathcal{L}$ inquiring a total of $n_e=\sum_{(X,Y) \in \mathcal{L}}|X||Y|$ matrix entries, \cref{alg:elem_extract} extracts all required elements in $\mathcal{O}(n_e\log n)$ operations. In other words, this algorithm requires $\mathcal{O}(\log n)$ operations per entry regardless of the number of entries needed. Consider for example the case where one wants to construct a butterfly matrix from the sum of two butterfly matrices. This can be done by calling BF\_entry\_eval with an extract routine, implemented using two calls to extract\_BF, which extracts entries from a given butterfly.
	
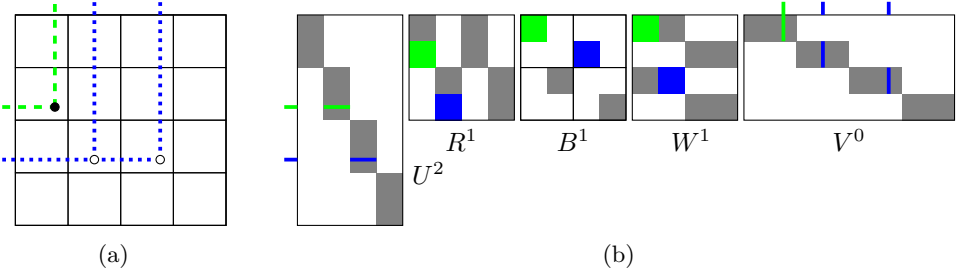
\begin{figure}
  \centering
  \begin{subfigure}[b]{.25\textwidth}
    \centering
    \begin{tikzpicture}[scale=2.8]
      \def\myshift{.03}
      \def\r{1/32}
      \def\rr{1/16}
      \def\h{1/2}
      \def\q{1/4}
      \def\o{1/8}
      \draw (0,0) rectangle (1,1);
      \foreach \x in {0,1,...,3} {
        \draw (\x*\q,0) rectangle ++(\q,1);
        \draw (0,\x*\q) rectangle ++(1,\q);
      }
      \draw [line width=.5mm,dashed,color=green] (-1/16,9/16) -- (3/16,9/16);
      \draw [line width=.5mm,dashed,color=green] (3/16,9/16) -- (3/16,1+1/16);
      \draw [fill=black] (3/16,9/16) circle (.02);

      \draw [line width=.5mm,dotted,color=blue] (-1/16,5/16) -- (11/16,5/16);
      \draw [line width=.5mm,dotted,color=blue] (6/16,5/16) -- (6/16,1+1/16);
      \draw [line width=.5mm,dotted,color=blue] (11/16,5/16) -- (11/16,1+1/16);
      \draw [fill=white] (6/16,5/16) circle (.02);
      \draw [fill=white] (11/16,5/16) circle (.02);

    \end{tikzpicture}
    \caption{}
  \end{subfigure}
  \hspace{.3cm}
  \begin{subfigure}[b]{.7\textwidth}
    \centering
    \begin{tikzpicture}[scale=2.8]
      \def\myshift{.53}
      \def\r{1/8}
      \def\rr{1/4}
      \def\h{1/2}
      \def\q{1/4}
      \def\o{1/8}
      \foreach \x in {0,1,...,3} {
        \fill [gray] (\x*\r,1-\x*\rr) rectangle ++(\r,-\rr);
      };
      \draw (0,0) rectangle (\h,1.0);
      \draw [line width=.5mm,color=blue] (-1/16,5/16) -- (0,5/16);
      \draw [line width=.5mm,color=blue] (2*\r,5/16) -- (3*\r,5/16);
      \draw [line width=.5mm,color=green] (-1/16,9/16) -- (0,9/16);
      \draw [line width=.5mm,color=green] (\r,9/16) -- (2*\r,9/16);
      \node[right] at (\h,\rr){$U^2$};

      \tikzset{shift={(\myshift,0)}}
      \foreach \x in {0,1} {
        \fill [gray] (\x*\r,1-\x*\rr) rectangle ++(\r,-\rr);
        \fill [gray] (\q+\x*\r,1-\x*\rr) rectangle ++(\r,-\rr);
      };
      \fill [green] (0,1-\r) rectangle ++(\r,-\r);
      \fill [blue] (\r,1-3*\r) rectangle ++(\r,-\r);
      \draw (0,1) rectangle (\h,\h);
      \node[below] at (\rr,\h){$R^1$};
      \tikzset{shift={(\myshift,0)}}
      \fill [green] (0,1) rectangle ++(\r,-\r);
      \fill [blue] (\q,1-\r) rectangle ++(\r,-\r);
      \fill [gray] (\r,1-\rr) rectangle ++(\r,-\r);
      \fill [gray] (\q+\r,1-\q-\r) rectangle ++(\r,-\r);
      \foreach \x in {0,1} {
        \foreach \y in {0,1} {
          \draw (\x*\q,1-\y*\q) rectangle ++(\q,-\q);
        }
      }
      \draw (0,1) rectangle (\h,\h);
      \node[below] at (\rr,\h){$B^1$};
      \tikzset{shift={(\myshift,0)}}
      \foreach \x in {0,1} {
        \fill [gray] (\x*\rr,1-\x*\r) rectangle ++(\rr,-\r);
        \fill [gray] (\x*\rr,1-\q-\x*\r) rectangle ++(\rr,-\r);
      };
      \fill [green] (0,1) rectangle ++(\r,-\r);
      \fill [blue] (\r,1-2*\r) rectangle ++(\r,-\r);
      \draw (0,1) rectangle (\h,\h);
      \tikzset{shift={(\myshift,0)}}
      \foreach \x in {0,1,...,3} {
        \fill [gray] (\x*\rr,1-\x*\r) rectangle ++(\rr,-\r);
      };
      \draw (0,1.0) rectangle (1,\h);
      \node[below] at (-\rr,\h){$W^1$};
      \draw [line width=.5mm,color=blue] (6/16,1) -- (6/16,1+1/16);
      \draw [line width=.5mm,color=blue] (11/16,1) -- (11/16,1+1/16);
      \draw [line width=.5mm,color=blue] (6/16,1-\r) -- (6/16,1-2*\r);
      \draw [line width=.5mm,color=blue] (11/16,1-2*\r) -- (11/16,1-3*\r);

      \draw [line width=.5mm,color=green] (3/16,1) -- (3/16,1+1/16);
      \draw [line width=.5mm,color=green] (3/16,1) -- (3/16,1-\r);
      
       \node[below] at (2*\rr,\h){$V^0$};       
    \end{tikzpicture}
    \caption{\label{fig:extract_BF_decomp}}
  \end{subfigure}
    \vspace{-15pt}
  \caption{The extract routine, see \cref{alg:elem_extract},
    to compute a list of submatrices from a 2-level butterfly matrix. (a) This shows the center level partitioning of the 2-level butterfly matrix and the two submatrices (with sizes $1\times 1$ and $1\times 2$, colored green and blue respectively) to be extracted. (b) The transfer, interpolation and skeleton matrices required for the extraction of the two subblocks are highlighted. \label{fig:extract}}
\vspace{-20pt}
\end{figure}

	In a nutshell, extracting a submatrix from a butterfly can be \ylrev{viewed as the product of three matrices $EAF$} with selection matrices $E$ and $F$ that pick the rows and columns of the submatrix. However, an efficient algorithm requires multiplying only selected transfer, interpolation and skeleton matrices. Specifically, \cref{alg:elem_extract} computes, for each $(X,Y)\in \mathcal{L}$, lists $\mathcal{L}_l$ of $(\tau,\nu)$ pairs indicating the required butterfly blocks (see line \ref{line:listl}). These blocks are then multiplied together to compute the submatrix $K(X,Y)$ (see lines \ref{line:ER}, \ref{line:WF}, \ref{line:EBF}), which requires $\mathcal{O}(n_e\log n)$ operations. For example, \cref{fig:extract} shows an extraction of two submatrices (with sizes $1\times 1$ and $1\times 2$, colored green and blue) from a 2-level butterfly, with the required transfer, interpolation and skeleton matrices also highlighted. To further improve the performance, we modify \cref{alg:elem_extract} by moving the outermost loop into the innermost loops at lines \ref{line:loopl} and \ref{line:loopr}. This way any butterfly block is multiplied at most once, and the communication is minimized when $A$ is distributed over multiple processes.    
	
	\section{Hierarchically Off-Diagonal Butterfly Matrix Representation\label{sec:HODBF}}
	
	The hierarchically off-diagonal low-rank (HOD-LR) matrix representation is a special case of the more general class of $\mathcal{H}$ matrices. For HOD-LR every off-diagonal block is assumed to be low-rank, which corresponds to so-called $\mathcal{H}$-matrix \ylrev{with weak admissibility condition}~\cite{hackbusch2004hierarchical}. The hierarchically off-diagonal butterfly (HOD-BF) format, however, is a generalization of HOD-LR where low-rank approximation is replaced by butterfly decomposition \cite{Liu_2017_HODBF}.
	
	For dense linear systems arising from high-frequency wave equations, the HOD-BF format is a suitable matrix representation, since butterfly compression applied to the off-diagonal blocks reduces storage and solution complexity, as opposed to $\mathcal{H}$ or HOD-LR matrices which do not reduce complexity for such problems. The HOD-BF matrix format was first developed to solve 2D high-frequency Helmholtz equations with $\mathcal{O}(n\log^2n)$ memory and $\mathcal{O}(n^{3/2}\log n)$ time \cite{Liu_2017_HODBF}. Recent work shows that the same complexity can also be obtained for 3D Helmholtz equations despite the non-constant butterfly rank due to \ylrev{the weak admissibility condition}\cite{liu2020butterfly}. It is also worth mentioning that compared to butterfly-based $\mathcal{H}$ matrix compression \ylrev{with strong admissibility condition} \cite{guo2017butterfly,Han_2018_butterflyLUDielectric}, HOD-BF enjoys simpler butterfly arithmetic, smaller leading constants in complexity, and significantly better parallelization performance. In what follows, we briefly describe the HOD-BF format, which is used in \cref{sec:preconditioner} to construct the quasi-linear complexity multifrontal solver. 
	
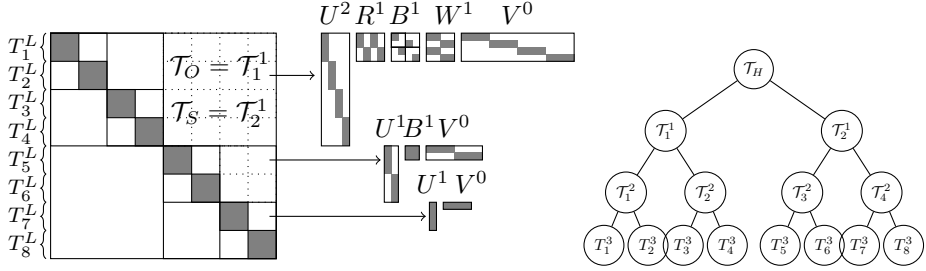
\begin{figure}
  \begin{center}
    \begin{tikzpicture}[scale=3]
      \def\myshift{.03}
      \def\r{1/32}
      \def\rr{1/16}
      \def\h{1/2}
      \def\q{1/4}
      \def\o{1/8}
      
      \draw [decorate,decoration={brace,amplitude=2pt},yshift=0pt,xshift=0cm] (-.02,1-0.125) -- (-.02,1-0.125*0) node [black,midway,xshift=-.3cm] {\footnotesize $T^L_1$};
      \draw [decorate,decoration={brace,amplitude=2pt},yshift=0pt,xshift=0cm] (-.02,1-0.125*2) -- (-.02,1-0.125*1) node [black,midway,xshift=-.3cm] {\footnotesize $T^L_2$};
      \draw [decorate,decoration={brace,amplitude=2pt},yshift=0pt,xshift=0cm] (-.02,1-0.125*3) -- (-.02,1-0.125*2) node [black,midway,xshift=-.3cm] {\footnotesize $T^L_3$};
      \draw [decorate,decoration={brace,amplitude=2pt},yshift=0pt,xshift=0cm] (-.02,1-0.125*4) -- (-.02,1-0.125*3) node [black,midway,xshift=-.3cm] {\footnotesize $T^L_4$};
      \draw [decorate,decoration={brace,amplitude=2pt},yshift=0pt,xshift=0cm] (-.02,1-0.125*5) -- (-.02,1-0.125*4) node [black,midway,xshift=-.3cm] {\footnotesize $T^L_5$};
      \draw [decorate,decoration={brace,amplitude=2pt},yshift=0pt,xshift=0cm] (-.02,1-0.125*6) -- (-.02,1-0.125*5) node [black,midway,xshift=-.3cm] {\footnotesize $T^L_6$};
      \draw [decorate,decoration={brace,amplitude=2pt},yshift=0pt,xshift=0cm] (-.02,1-0.125*7) -- (-.02,1-0.125*6) node [black,midway,xshift=-.3cm] {\footnotesize $T^L_7$};
      \draw [decorate,decoration={brace,amplitude=2pt},yshift=0pt,xshift=0cm] (-.02,1-0.125*8) -- (-.02,1-0.125*7) node [black,midway,xshift=-.3cm] {\footnotesize $T^L_8$};

      \draw (0,0) rectangle (1,1);
      \foreach \x in {0,1,...,7} {
        \draw [fill=gray] (\x*\o,1-\x*\o) rectangle ++(\o,-\o);
      }
      \foreach \x in {0,1,...,3} {
        \draw (\x*\q,1-\x*\q) rectangle ++(\q,-\q);
      }
      \foreach \x in {0,1} {
        \draw (\x*\h,1-\x*\h) rectangle ++(\h,-\h);
      }
      \foreach \x in {0,2} {
        \draw[dotted] (\h+\x*\o,\h) rectangle ++(\o,\h);
      }
      \foreach \x in {0,2} {
        \draw[dotted] (\h,1-\x*\o) rectangle ++(\h,-\o);
      }
      \node[above] at (\h+\q,\h+\q){$\mathcal{T}_O = \mathcal{T}^1_1$};
      \node[below] at (\h+\q,\h+\q){$\mathcal{T}_S = \mathcal{T}^1_2$};


      \draw[dotted] (\h+\q,\h) rectangle ++(\o,-\o);
      \draw[dotted] (1-\o,\q) rectangle ++(\o,\o);

      \draw [->] (1-\myshift,1-\q+\o / 2) -- ++(.2,0);
      \draw [->] (1-\myshift,\q+3./2.*\o) -- ++(.4+2*\r+\myshift,0);
      \draw [->] (1-\myshift,\o+\o/2) -- ++(.6+2*\r+\myshift,0);

      \tikzset{shift={(1.2,0)}}
      \foreach \x in {0,1,...,3} {
        \fill [gray] (\x*\r,1-\x*\o) rectangle ++(\r,-\o);
      };
      \draw (0,1) rectangle (\r*4,\h);
      \node[above] at (\r*2,1){$U^2$};
      \tikzset{shift={(\r*4+\myshift,0)}}
      \foreach \x in {0,1} {
        \fill [gray] (\x*\r,1-\x*\rr) rectangle ++(\r,-\rr);
        \fill [gray] (\o/2+\x*\r,1-\x*\rr) rectangle ++(\r,-\rr);
      };
      \draw (0,1) rectangle (\o,1-\o);
      \node[above] at (\o/2,1){$R^1$};
      \tikzset{shift={(\r*4+\myshift,0)}}
      \foreach \x in {0,1} {
        \foreach \y in {0,1} {
          \fill [gray] (\x*\o/2+\y*\r,1-\y*\o/2-\x*\r) rectangle ++(\r,-\r);
          \draw (\x*\o/2,1-\y*\o/2) rectangle ++(\o/2,-\o/2);
        }
      }
      \draw (0,1) rectangle (\o/2,1-\o/2);
      \node[above] at (\o/2,1){$B^1$};
      \tikzset{shift={(\r*4+\myshift,0)}}
      \foreach \x in {0,1} {
        \fill [gray] (\x*\rr,1-\x*\r) rectangle ++(\rr,-\r);
        \fill [gray] (\x*\rr,1-\x*\r-\o/2) rectangle ++(\rr,-\r);
      };
      \draw (0,1) rectangle (\o,1-\o);
      \node[above] at (\o,1){$W^1$};
      \tikzset{shift={(\r*4+\myshift,0)}}
      \foreach \x in {0,1,...,3} {
        \fill [gray] (\x*\o,1-\x*\r) rectangle ++(\o,-\r);
      };
      \draw (0,1) rectangle (\h,1-\o);
      \node[above] at (\q,1){$V^0$};

      \tikzset{shift={(-9*\r-2*\myshift,0)}}
      \foreach \x in {0,1} {
        \fill [gray] (\x*\r,\h-\x*\o) rectangle ++(\r,-\o);
      };
      \draw (0,\q) rectangle (\r*2,\h);
      \node[above] at (\r,\h){$U^1$};
      
      \tikzset{shift={(\r*2+\myshift,0)}}
      \fill [gray] (0,\h) rectangle (\o/2,\h-\o/2);
      \draw (0,\h) rectangle (\o/2,\h-\o/2);
      \node[above] at (\o/2,\h){$B^1$};

      \tikzset{shift={(\r*2+\myshift,0)}}
      \foreach \x in {0,1} {
        \fill [gray] (\x*\o,\h-\x*\r) rectangle ++(\o,-\r);
      };
      \draw (0,\h) rectangle (\q,\h-\o/2);
      \node[above] at (\o,\h){$V^0$};
      
      \tikzset{shift={(-4*\r-2*\myshift+.2,0)}}
      \fill [gray] (0,\q) rectangle (\r,\q-\o);
      \draw (0,\q) rectangle (\r,\q-\o);
      \node[above] at (\r/2,\q){$U^1$};
      
      \tikzset{shift={(\r+\myshift,0)}}
      \fill [gray] (0,\q) rectangle (\o,\q-\r);
      \draw (0,\q) rectangle (\o,\q-\r);
      \node[above] at (\o,\q){$V^0$};
    \end{tikzpicture}
    \begin{tikzpicture}[scale=.18]
      \node[cnode](14){$\mathcal{T}_H$}
      child{node[cnode](6){$\mathcal{T}^1_1$}
        child{node[cnode](2){$\mathcal{T}^2_1$}
                child{node[cnode](0){$T^{3}_1$}} child{node[cnode](1){$T^{3}_2$}} }
        child{node[cnode](5){$\mathcal{T}^2_2$}
                child{node[cnode](3){$T^3_3$}} child{node[cnode](4){$T^3_4$}} }
        }
      child{node[cnode](13){$\mathcal{T}^1_2$}
        child{node[cnode](9){$\mathcal{T}^2_3$}   
                 child{node[cnode](7){$T^3_5$}}   child{node[cnode](8){$T^3_6$}} }
        child{node[cnode](12){$\mathcal{T}^2_4$} 
                  child{node[cnode](10){$T^3_7$}} child{node[cnode](11){$T^3_8$}} }
      };
    \end{tikzpicture}

  \end{center}
 \caption{Illustration of a $4$-level hierarchically off-diagonal butterfly matrix. The root node is at level $l = 0$, all the leaf nodes are at level $L=3$. The two largest off-diagonal blocks are approximated using $2$-level butterfly matrices. The $4$ off-diagonal blocks one level down in the hierarchy are approximated using a $1$ level butterfly ($U^1B^1V^0$). Finally, the smallest off-diagonal blocks are approximated as low-rank, i.e., $0$-level butterfly matrices. Note that these different butterfly blocks are not related. The hierarchy is illustrated using the tree on the right. Each leaf node stores a dense diagonal block $D_\tau$, the parent nodes store $2$ off-diagonal (butterfly) blocks.\label{fig:HOD-BF}}
\end{figure}

	As illustrated in \cref{fig:HOD-BF}, in the HOD-BF format diagonal blocks are recursively refined until a certain minimum size is reached. For a square matrix $A \in \mathbb{R}^{n \times n}$, this partitioning defines a single binary tree $\mathcal{T}_H$, as shown on the right in \cref{fig:HOD-BF}. The root node is at level $0$; its children are at level $1$, etc.  All the leaf nodes are at level $L$. Each node $\tau$ at level $l$ in the HOD-BF tree has an index set $T^l_\tau \subset T_H = \{1,\dots,n\}$, where $T_H$ is the index set corresponding to all rows and columns of the matrix. For an internal node $\tau$ at level $l$ with children $\tau_1$ and $\tau_2$, $T_\tau^l = T_{\tau_1}^{l+1} \cup T_{\tau_2}^{l+1}$. At the lowest level of the hierarchy, the leaves of the HOD-BF tree, the diagonal blocks $D_\tau = A(T^L_\tau, T^L_\tau)$ are stored as regular dense matrices, while off-diagonal blocks are approximated using butterfly decomposition. Let $\tau_1$ and $\tau_2$ be two siblings in $\mathcal{T}_H$ on level $l$ with the two trees $\mathcal{T}^l_{\tau_1}$ and $\mathcal{T}^l_{\tau_2}$, subtrees of $\mathcal{T}_H$, rooted at nodes $\tau_1$ and $\tau_2$ respectively. These two sibling nodes correspond to two off-diagonal blocks $B_{\tau_1} = A(T^l_{\tau_1}, T^l_{\tau_2})$ and $B_{\tau_2} = A(T^l_{\tau_2}, T^l_{\tau_1})$, approximated using butterfly decomposition. One of those butterfly blocks is defined by $\mathcal{T}_O = \mathcal{T}^l_{\tau_1}$ and $\mathcal{T}_S = \mathcal{T}^l_{\tau_2}$, while the other is defined by $\mathcal{T}_O = \mathcal{T}^l_{\tau_2}$ and $\mathcal{T}_S = \mathcal{T}^l_{\tau_1}$.

	
	
	\subsection{HOD-BF Construction Using Entry Evaluation\label{ssec:HODB_entry_eval}}
	An HOD-BF matrix representation based on sampling matrix entries can be constructed upon applying the BF\_entry\_eval algorithm (\cref{alg:BF_entry}) to all off-diagonal blocks of the HOD-BF matrix. The construction can be done in $\mathcal{O}(n\log^2n)$, or in $\mathcal{O}(n\log^3n)$ operations, if an individual matrix entry can be computed in $\mathcal{O}(1)$, or in $\mathcal{O}(\log n)$ time. We name the HOD-BF construction of a matrix $A$ as HODBF\_entry\_eval($A$), where $A$ is passed in the form of a routine that extracts a list of (rows, columns) index sets from $A$.
	
	Similar to the butterfly extract routine in \cref{ssec::elem-extraction}, we also implement a routine to extract a list $\mathcal{L}$ of (rows, columns) index sets from an HOD-BF matrix $A$, called extract\_HODBF($\mathcal{L}, A$). This routine is implemented using extract\_BF for the off-diagonal blocks of $A$.

	\subsection{Inversion of HOD-BF Matrices\label{ssec:HODBFinversion}}
	Once constructed, the inverse of the HOD-BF matrix can be computed in $\mathcal{O}(n^{3/2}\log n)$ operations based on the randomized matrix-vector product algorithm BF\_random\_matvec described in \cref{ssec::random-bf-construction}. The inversion algorithm has been previously described in~\cite{Liu_2017_HODBF} and is briefly summarized as HODBF\_invert, \cref{alg:hodbf_invert}.
	
	Let $D_\tau=A$ with $\tau$ denoting the root node of $\mathcal{T}_H$. The algorithm first computes $D_{\tau_1}^{-1}$ and $D_{\tau_2}^{-1}$ using two recursive calls. Then the two off-diagonal butterflies are updated as $B_{\tau_i}\leftarrow D^{-1}_{\tau_i}B_{\tau_i}$ using BF\_random\_matvec (at lines \ref{line:matvec1_hodbf} and \ref{line:matvec2_hodbf}) as both $D^{-1}_{\tau_i}$ and $B_{\tau_i}$ are already compressed. Finally the updated matrix $[I,B_{\tau_1};B_{\tau_2},I]$ is inverted using the butterfly extension of the Sherman-Morrison-Woodbury formula \cite{Hager1989SMW}, named BF\_SMW, which in turn requires BF\_random\_matvec (at lines \ref{line:matvec3_hodbf} and \ref{line:matvec4_hodbf}) to facilitate the computation.

	\begin{algorithm}
		\caption{${\rm HODBF\_invert}(A)$: Inversion of a square HOD-BF matrix.}
		\label{alg:hodbf_invert}
		\hspace*{\algorithmicindent} \textbf{Input:} $A$ in HOD-BF form with $L$ levels \\
		\hspace*{\algorithmicindent} \textbf{Output:} $A^{-1}$ in HOD-BF form 	
		\begin{algorithmic}[1]
			\State Let $D_{\tau}=A$ with $\tau$ denoting the root node.
			\If{$D_{\tau}$ dense}
			\State Directly compute $D_\tau^{-1}$
			\Else
			\State $D_{\tau_1}^{-1}\leftarrow$ HODBF\_invert$\left( D_{\tau_1} \right)$ \Comment{$D_{\tau_1}$ is HODBF with $L-1$ levels}
			\State $D_{\tau_2}^{-1}\leftarrow$ HODBF\_invert$\left( D_{\tau_2} \right)$ \Comment{$D_{\tau_2}$ is HODBF with $L-1$ levels}
			\State $B_{\tau_1}\leftarrow$ BF\_random\_matvec$\left( D_{\tau_1}^{-1}B_{\tau_1} \right)$\label{line:matvec1_hodbf}
			\State $B_{\tau_2}\leftarrow$ BF\_random\_matvec$\left( D_{\tau_2}^{-1}B_{\tau_2} \right)$\label{line:matvec2_hodbf}
			\State $D_{\tau}^{-1}\leftarrow$ BF\_SMW$\left( \begin{bmatrix}
			I& B_{\tau_1} \\
			B_{\tau_2}&I 
			\end{bmatrix} \right) $$\begin{bmatrix}D_{\tau_1}^{-1}&\\&D_{\tau_2}^{-1}\end{bmatrix}$
			\EndIf
			\Function {BF\_SMW}{$A$}
			\State \textbf{Input:} $A-I$ is a butterfly of $L$ levels \Comment{If $L=0$, the low-rank SMW \cite{Hager1989SMW} can be used instead.}
			\State \textbf{Output:} $A^{-1}$ as a butterfly of $L$ levels added with the identity $I$  
			\State Split $A$ into four children butterflies of $L-2$ levels: $A=[A_{11},A_{12};A_{21},A_{22}]$ using $\mathcal{T}_O$ and $\mathcal{T}_S$
			\State $A_{22}^{-1}\leftarrow$ BF\_SMW$\left( A_{22} \right)$
			\State $A_{11}\leftarrow$ BF\_random\_matvec$\left( A_{11}-A_{12}(I+A_{22})A_{21}\right)$\label{line:matvec3_hodbf}
			\State $A_{11}^{-1}\leftarrow$ BF\_SMW$\left(A_{11}\right)$
			\State $A^{-1}\!\leftarrow$\! $I+$BF\_random\_matvec$ \left(\! \begin{bmatrix}
			I & \\
			-A_{22}^{-1}A_{21} & I
			\end{bmatrix}\!\!\!\begin{bmatrix}
			A_{11}^{-1} & \\
			& A_{22}^{-1}
			\end{bmatrix}\!\!\!\begin{bmatrix}
			I &-A_{12}A_{22}^{-1} \\
			& I
			\end{bmatrix}\!\!-\!I \!\right)$\label{line:matvec4_hodbf}
			\EndFunction		
		\end{algorithmic}
		
	\end{algorithm}

	\section{Rank Structured Multifrontal Factorization
		\label{sec:preconditioner}}
	
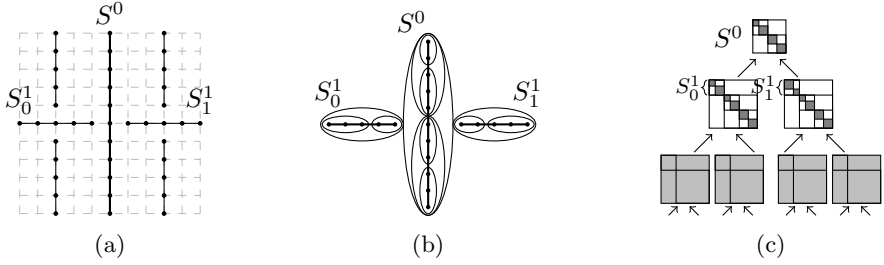
\begin{figure}
  \centering
  \begin{subfigure}[b]{.3\textwidth}
    \centering
    \begin{tikzpicture}[scale=1.2]
      \def\l{2}
      \def\h{1}
      \def\d{.2}
      \foreach \z in {0,...,10} {
        \draw[lightgray,dashed] (\z * \d,0) -- (\z * \d, \l);
        \draw[lightgray,dashed] (0,\z * \d) -- (\l, \z * \d);
      }
      \node[above] at (\h,\l){$S^0$};

      \node[above] at (0,\h){$S^1_0$};
      \node[above] at (\l,\h){$S^1_1$};


      \draw[thick] (\h,0) -- (\h,\l);
      \foreach \z in {0,...,10} {
         \draw[fill] (\h,\z * \d) circle (0.02);
      }
      \draw (0,\h) -- (4*\d,\h);
      \draw (6*\d,\h) -- (\l,\h);
      \foreach \z in {0,...,4} {
         \draw[fill] (\z * \d,\h) circle (0.02);
         \draw[fill] (\h + \d + \z * \d,\h) circle (0.02);
      }

      \draw (2*\d,\h+\d) -- (2*\d,\l);
      \draw (2*\d,0) -- (2*\d,\h - \d);
      \draw (8*\d,\h+\d) -- (8*\d,\l);
      \draw (8*\d,0) -- (8*\d,\h - \d);
      \foreach \z in {0,...,4} {
         \draw[fill] (2*\d,\z * \d) circle (0.02);
         \draw[fill] (2*\d,\z * \d + \h +\d) circle (0.02);

         \draw[fill] (8*\d,\z * \d) circle (0.02);
         \draw[fill] (8*\d,\z * \d + \h +\d) circle (0.02);
      }
    \end{tikzpicture}
    \caption{\label{fig::mesh_11}}
   \end{subfigure}
   \hfill
   \begin{subfigure}[b]{.3\textwidth}
     \centering
      \begin{tikzpicture}[scale=1.1]
      \def\l{2}
      \def\h{1}
      \def\d{.2}
      \node[above] at (\h,\l){$S^0$};
      \node[above] at (0,\h+\d/2){$S^1_0$};
      \node[above] at (\l+2 * \d,\h+\d/2){$S^1_1$};
      \draw[thick] (\d+ \h,0) -- (\d+ \h,\l);
      \foreach \z in {0,...,10} {
         \draw[fill] (\d+ \h,\z * \d) circle (0.02);
      }
      \draw[thick] (0,\h) -- (\h - \d,\h);
      \draw[thick] (2*\d+ \h + \d,\h) -- (2*\d+ \l,\h);
      \foreach \z in {0,...,4} {
         \draw[fill] (\d * \z,\h) circle (0.02);
         \draw[fill] (8*\d + \d * \z,\h) circle (0.02);
      }
      
      \draw (\d+ \h,5 * \d) circle [x radius=1.5 * \d, y radius=5.5 *\d];
      \draw (\d+ \h,2.5 * \d) circle [x radius=\d, y radius=3 *\d];
      \draw (\d+ \h,1 * \d) circle [x radius=\d / 2, y radius=1.4 *\d];
      \draw (\d+ \h,4 * \d) circle [x radius=\d / 2, y radius=1.4 *\d];
      \draw (\d+ \h,8 * \d) circle [x radius=\d, y radius=2.5 *\d];
      \draw (\d+ \h,7 * \d) circle [x radius=\d / 2, y radius=1.4 *\d];
      \draw (\d+ \h,9.5 * \d) circle [x radius=\d / 2, y radius=.9 *\d];
      
      \draw (\d,\h) circle [x radius=1.4 * \d, y radius=\d / 2];
      \draw (3.5 * \d,\h) circle [x radius=.9 * \d, y radius=\d / 2];
      \draw (2 * \d,\h) circle [x radius=2.5 * \d, y radius=\d];

      \draw (11*\d,\h) circle [x radius=1.4 * \d, y radius=\d / 2];
      \draw (8.5 * \d,\h) circle [x radius=.9 * \d, y radius=\d / 2];
      \draw (10 * \d,\h) circle [x radius=2.5 * \d, y radius=\d];
    \end{tikzpicture}
    \caption{\label{fig:sep_order}}
  \end{subfigure}
  \hfill
  \begin{subfigure}[b]{.35\textwidth}
  \centering
  \begin{tikzpicture}[scale=.4]
    \def\r{1/32}
    \def\rr{1/16}
    \def\h{1/2}
    \def\q{1/4}
    \def\o{1/8}
    \draw (0,0) rectangle (1.1,1.1);
    \node[left] at (0,0.55){$S^0$};
    \draw [fill=gray] (0,1.1) rectangle ++(.2,-.2);
    \draw [fill=gray] (.2,.9) rectangle ++(.3,-.3);
    \draw [fill=gray] (.5,.6) rectangle ++(.3,-.3);
    \draw [fill=gray] (.8,.3) rectangle ++(.3,-.3);
    \draw (0,1.1) rectangle ++(.5,-.5);
    \draw (.5,.6) rectangle ++(.6,-.6);

    \draw [->] (-.5,-.8) -- ++(.6,.6);
    \draw [->] (1.5,-.8) -- ++(-.6,.6);

    \tikzset{shift={(-1.45,-2.5)}}
    \draw [decorate,decoration={brace,amplitude=2pt},yshift=0pt,xshift=0cm] (-.07,1.1) -- (-.07,1.1+0.5) node [black,midway,xshift=-.25cm] {\footnotesize $S^1_0$};

    \draw (0,0) rectangle (.5+1.1,.5+1.1);
    \draw (0,1.1) rectangle (.5,1.1+.5);
    \draw (0,1.1) rectangle (.5,1.1+.5);
    \draw [fill=gray] (0,1.6) rectangle ++(.2,-.2);
    \draw [fill=gray] (.2,1.4) rectangle ++(.3,-.3);

    \draw [->] (-.5,-.8) -- ++(.6,.6);
    \draw [->] (1.5,-.8) -- ++(-.6,.6);

    \tikzset{shift={(.5,0)}}
    \draw (0,0) rectangle (1.1,1.1);
    \draw [fill=gray] (0,1.1) rectangle ++(.2,-.2);
    \draw [fill=gray] (.2,.9) rectangle ++(.3,-.3);
    \draw [fill=gray] (.5,.6) rectangle ++(.3,-.3);
    \draw [fill=gray] (.8,.3) rectangle ++(.3,-.3);
    \draw (0,1.1) rectangle ++(.5,-.5);
    \draw (.5,.6) rectangle ++(.6,-.6);

    \tikzset{shift={(2,0)}}
    \draw [decorate,decoration={brace,amplitude=2pt},yshift=0pt,xshift=0cm] (-.07,1.1) -- (-.07,1.1+0.5) node [black,midway,xshift=-.25cm] {\footnotesize $S^1_1$};
    \draw (0,0) rectangle (.5+1.1,.5+1.1);
    \draw (0,1.1) rectangle (.5,1.1+.5);
    \draw (0,1.1) rectangle (.5,1.1+.5);
    \draw [fill=gray] (0,1.6) rectangle ++(.3,-.3);
    \draw [fill=gray] (.3,1.3) rectangle ++(.2,-.2);

    \tikzset{shift={(.5,0)}}
    \draw (0,0) rectangle (1.1,1.1);
    \draw [fill=gray] (0,1.1) rectangle ++(.2,-.2);
    \draw [fill=gray] (.2,.9) rectangle ++(.3,-.3);
    \draw [fill=gray] (.5,.6) rectangle ++(.3,-.3);
    \draw [fill=gray] (.8,.3) rectangle ++(.3,-.3);
    \draw (0,1.1) rectangle ++(.5,-.5);
    \draw (.5,.6) rectangle ++(.6,-.6);

    \draw [->] (-.5,-.8) -- ++(.6,.6);
    \draw [->] (1.5,-.8) -- ++(-.6,.6);

    \tikzset{shift={(-4.6,-2.5)}}
    \draw [fill=lightgray] (0,0) rectangle (.5+1.1,.5+1.1);
    \draw (0,.5+1.1) rectangle (.5,1.1);
    \draw (.5,1.1) rectangle (.5+1.1,0);
    \draw [->] (.25,-.4) -- ++(.3,.3);
    \draw [->] (1.25,-.4) -- ++(-.3,.3);

    \tikzset{shift={(1.8,0)}}
    \draw [fill=lightgray] (0,0) rectangle (.5+1.1,.5+1.1);
    \draw (0,.5+1.1) rectangle (.5,1.1);
    \draw (.5,1.1) rectangle (.5+1.1,0);
    \draw [->] (.25,-.4) -- ++(.3,.3);
    \draw [->] (1.25,-.4) -- ++(-.3,.3);

    \tikzset{shift={(2.1,0)}}
    \draw [fill=lightgray] (0,0) rectangle (.5+1.1,.5+1.1);
    \draw (0,.5+1.1) rectangle (.5,1.1);
    \draw (.5,1.1) rectangle (.5+1.1,0);
    \draw [->] (.25,-.4) -- ++(.3,.3);
    \draw [->] (1.25,-.4) -- ++(-.3,.3);

    \tikzset{shift={(1.8,0)}}
    \draw [fill=lightgray] (0,0) rectangle (.5+1.1,.5+1.1);
    \draw (0,.5+1.1) rectangle (.5,1.1);
    \draw (.5,1.1) rectangle (.5+1.1,0);
    \draw [->] (.25,-.4) -- ++(.3,.3);
    \draw [->] (1.25,-.4) -- ++(-.3,.3);
  \end{tikzpicture}
  \caption{\label{fig:MF_BF_tree}}
  \end{subfigure}
  \caption{ (a) The top three levels of nested dissection for an $11^2$ mesh. (b) The root separator $S^0$ is a vertical $11$ point line, which is recursively bisected to define the hierarchical matrix partitioning. The next level separators $S^1_0$ and $S^1_1$, are similarly partitioned. (c) The root separator corresponds to the top level front, and its HOD-BF partitioning is defined by the recursive bisection of the root separator, as shown in (b), and similarly for the next level down in the assembly/frontal tree. For the lower levels, the fronts are regular dense matrices. Note that the fronts in (c) are to scale, but from this figure it is not obvious that the fronts typically get smaller lower in the tree (except for the root front, which has no Schur complement). Only the top 3 fronts are compressed using HOD-BF, while the others are treated as regular dense matrices. \label{fig:separator_ordering}}
\end{figure}

	It has been studied by several authors that although the frontal matrices are dense, they are data-sparse for many applications and can often be well approximated using rank-structured matrix formats. \cref{alg:preconditioner} outlines the rank-structured multifrontal factorization using HOD-BF compression for the fronts. However since the more complicated HOD-BF matrix format has overhead for smaller matrices -- compared to the highly optimized BLAS and LAPACK routines -- HOD-BF compression is only used for fronts larger than a certain threshold $n_{\text{min}}$. Typically, the larger fronts are found closer to the root of the multifrontal assembly tree. This is illustrated in \cref{fig:separator_ordering} for a
	small regular \ylrev{$5\times5\times4$} mesh (\cref{fig::mesh_11}), and
	\cref{fig:MF_BF_tree} shows the corresponding multifrontal assembly tree,
	where only the top three fronts are compressed using HOD-BF.
	
	We now discuss the construction and partial factorization of the HOD-BF compressed fronts. To limit the overall complexity of the solver, a large front in the rank-structured multifrontal solver is never explicitly assembled fully as a large dense matrix. Instead, the solver relies on butterfly and HOD-BF construction using either element extraction, as described in \cref{ssec::extraction-bf-construction,sec:HODBF} or randomized sampling, as in \cref{ssec::random-bf-construction}. Recall that a front $F_\tau$ is built up from elements of the reordered sparse input matrix $A$, and \ylrev{the contribution blocks of the children of the front in the assembly tree: $\ylrev{C}_{\nu_1}$ and $\ylrev{C}_{\nu_2}$}, where $\nu_1$ and $\nu_2$ are the two children of $\tau$. Since multifrontal factorization traverses the assembly tree from the leaves to the root, these children contribution blocks might already be compressed using the HOD-BF format. Hence, extracting frontal matrix elements requires getting them from fronts previously compressed as HOD-BF. The end result looks like:
	\begin{equation}
  F_\tau  = 
      \begin{tikzpicture}[baseline={([yshift=-.8ex]current bounding box.center)}]
        \def\l{2/3}
        \def\o{\l/8}
        \def\h{\l/2}
        \def\q{\l/4}
        \draw (0,2) rectangle (\l,2-\l);
        \foreach \x in {0,1,...,3} {
          \draw [fill=gray] (\x*\q,\l-\x*\q+4/3) rectangle ++(\q,-\q);
        }
        \foreach \x in {0,1} {
          \draw (\x*\h,\l-\x*\h+4/3) rectangle ++(\h,-\h);
        }
        \def\l{4/3}
        \def\o{\l/8}
        \def\h{\l/2}
        \def\q{\l/4}
        \draw (0,0) rectangle (\l/2,\l);
        \draw (\l/2,2) rectangle (2,\l);
        \node[auto] at (4/3,2-1/3){$F_{12}$};
        \node[auto] at (1/3,2/3){$F_{21}$};
        \draw (\l/2,2-\l/2) rectangle (2,0);
        \foreach \x in {0,1,...,7} {
          \draw [fill=gray] (\x*\o+2/3,\l-\x*\o) rectangle ++(\o,-\o);
        }
        \foreach \x in {0,1,...,3} {
          \draw (\x*\q+2/3,\l-\x*\q) rectangle ++(\q,-\q);
        }
        \foreach \x in {0,1} {
          \draw (\x*\h+2/3,\l-\x*\h) rectangle ++(\h,-\h);
        }
  \end{tikzpicture} =
    \begin{tikzpicture}[baseline={([yshift=-2.3ex]current bounding box.center)}]
        \def\l{2}
        \node[above] at (\l/2,\l){\emph{sparse}};
        \draw (0,0) rectangle (\l/3,\l-\l/3);
        \draw (0,\l-\l / 3) rectangle (\l / 3,\l);
        \draw (\l/3,\l-\l / 3) rectangle (\l,\l);
        \foreach \x in {0,1,2,...,9} {
          \draw [fill=black] (\x * \l / 30 + \l / 60,\l - \x * \l / 30 - \l / 60) circle (\l / 120);
        }
        \foreach \x in {0,1,...,8} {
          \draw [fill=black] (\x * \l / 30 + \l / 30 + \l / 60,\l - \x * \l / 30 - \l / 60) circle (\l / 120);
        }
        \foreach \x in {0,1,...,8} {
          \draw [fill=black] (\x * \l / 30 + \l / 60,\l - \x * \l / 30 - \l / 60 - \l / 30) circle (\l / 120);
        }
        \draw [fill=black] (\l/2,\l - \l / 8) circle (\l / 120);
        \draw [fill=black] (\l/8,\l - \l / 2) circle (\l / 120);
        \draw [fill=black] (2*\l/3,\l - \l / 6) circle (\l / 120);
        \draw [fill=black] (\l/6,\l - 2*\l / 3) circle (\l / 120);
      \end{tikzpicture}
      \extadd 
      \begin{tikzpicture}[baseline={([yshift=-2.5ex]current bounding box.center)}]
        \def\l{1.}
        \def\o{\l/8}
        \def\h{\l/2}
        \def\q{\l/4}
        \node[above] at (\l/2,\l){$\tiny C\!B_{\nu_1}$};
        \draw (0,0) rectangle (\l,\l);
        \foreach \x in {0,1,...,3} {
          \draw [fill=gray] (\x*\q,\l-\x*\q) rectangle ++(\q,-\q);
        }
        \foreach \x in {0,1} {
          \draw (\x*\h,\l-\x*\h) rectangle ++(\h,-\h);
        }
      \end{tikzpicture}
      \extadd
      \begin{tikzpicture}[baseline={([yshift=-2.3ex]current bounding box.center)}]
        \def\l{1.4}
        \def\o{\l/8}
        \def\h{\l/2}
        \def\q{\l/4}
        \node[above] at (\l/2,\l){$C\!B_{\nu_2}$};
        \draw (0,0) rectangle (\l,\l);
        \foreach \x in {0,1,...,7} {
          \draw [fill=gray] (\x*\o,\l-\x*\o) rectangle ++(\o,-\o);
        }
        \foreach \x in {0,1,...,3} {
          \draw (\x*\q,\l-\x*\q) rectangle ++(\q,-\q);
        }
        \foreach \x in {0,1} {
          \draw (\x*\h,\l-\x*\h) rectangle ++(\h,-\h);
        }
      \end{tikzpicture} \, , \label{eq:front_assembly}
\end{equation}
	with $F_{11}$ and $F_{22}$ 
	compressed as HOD-BF, and $F_{12}$ and $F_{21}$ compressed as butterfly. For each front to be compressed, the following operations are in order:
	\begin{enumerate}[leftmargin=*]
		\item At first, the $F_{11}$ block of $F \equiv F_\tau$ is compressed as an HOD-BF matrix via HODBF\_entry\_eval, see \cref{ssec:HODB_entry_eval}, which calls BF\_entry\_eval,  \cref{alg:BF_entry}, for each of the off-diagonal blocks, using a routine extract($\mathcal{L},F_{11}$) to extract elements from $F_{11} = A(I_{\tau}^{\text{s}},I_{\tau}^{\text{s}}) \extadd \ylrev{C}_{{\nu_1}} \extadd \ylrev{C}_{{\nu_2}}$, see line~\ref{line:F11_HODBF} in \cref{alg:preconditioner}. Here $\ylrev{C}_{{\nu_1}}$ refers to the contribution block, the $F_{\nu_1;22}$ block including its Schur update, of child $\nu_1$ of node $\tau$ in the assembly tree. Note that in this case, the extend-add operation just requires checking whether the required matrix entries appear in the sparse matrix, or in the child contribution blocks, and then adding those different contributions together. 
		Consider for example the extraction of a single $2 \times 2$ subblock from a front, i.e., $\mathcal{L} = \{ (\{x_1, x_2\}, \{y_1, y_2\})\}$ is a list with a single (rows, columns) index set. Note that in general, the list can contain multiple index sets for extracting multiple subblocks. This might look as follows:
		\begin{equation}
    \begin{tikzpicture}[baseline={([yshift=-.8ex]current bounding box.center)}]
        \def\l{2}
        \draw (0,0) rectangle (\l,\l);
        \draw [thick, densely dotted,color=blue] (-0.1,\l - 5 * \l / 30 - \l / 60) -- (5 * \l / 30 - \l / 60, \l - 5 * \l / 30 - \l / 60);
        \draw [thick, densely dotted,color=blue] (-0.1,\l - 8 * \l / 30 - \l / 60) -- (5 * \l / 30 - \l / 60, \l - 8 * \l / 30 - \l / 60);

        \draw [thick, densely dotted,color=blue] (2 * \l / 30 - \l / 60,\l - 8 * \l / 30 - \l / 60) -- (2 * \l / 30 - \l / 60,\l + 0.1);
        \draw [thick, densely dotted,color=blue] (5 * \l / 30 - \l / 60,\l - 8 * \l / 30 - \l / 60) -- (5 * \l / 30 - \l / 60,\l + 0.1);

        \node[left] at (-0.1,\l - 5 * \l / 30 - \l / 60){$x_1$};
        \node[left] at (-0.1,\l - 8 * \l / 30 - \l / 60){$x_2$};
        \node[above] at (2 * \l / 30 - \l / 60 - 0.05, \l + .05){$y_1$};
        \node[above] at (5 * \l / 30 - \l / 60 + 0.05, \l + .05){$y_2$};

        \draw [fill=white] (5 * \l / 30 - \l / 60, \l - 5 * \l / 30 - \l / 60) circle (\l / 60);
        \draw [fill=white] (2 * \l / 30 - \l / 60, \l - 5 * \l / 30 - \l / 60) circle (\l / 60);
        \draw [fill=white] (2 * \l / 30 - \l / 60, \l - 8 * \l / 30 - \l / 60) circle (\l / 60);
        \draw [fill=white] (5 * \l / 30 - \l / 60, \l - 8 * \l / 30 - \l / 60) circle (\l / 60);
    \end{tikzpicture} =
    \begin{tikzpicture}[baseline={([yshift=-.8ex]current bounding box.center)}]
        \def\l{2}
        \draw (0,0) rectangle (\l/3,\l-\l/3);
        \draw (0,\l-\l / 3) rectangle (\l / 3,\l);
        \draw (\l/3,\l-\l / 3) rectangle (\l,\l);
        
        \draw [thick, densely dotted,color=blue] (-0.1,\l - 5 * \l / 30 - \l / 60) -- (5 * \l / 30 - \l / 60, \l - 5 * \l / 30 - \l / 60);
        \draw [thick, densely dotted,color=blue] (-0.1,\l - 8 * \l / 30 - \l / 60) -- (5 * \l / 30 - \l / 60, \l - 8 * \l / 30 - \l / 60);

        \draw [thick, densely dotted,color=blue] (2 * \l / 30 - \l / 60,\l - 8 * \l / 30 - \l / 60) -- (2 * \l / 30 - \l / 60,\l + 0.1);
        \draw [thick, densely dotted,color=blue] (5 * \l / 30 - \l / 60,\l - 8 * \l / 30 - \l / 60) -- (5 * \l / 30 - \l / 60,\l + 0.1);
        
        \node[left] at (-0.1,\l - 5 * \l / 30 - \l / 60){$x_1$};
        \node[left] at (-0.1,\l - 8 * \l / 30 - \l / 60){$x_2$};

        \node[above] at (2 * \l / 30 - \l / 60 - 0.05, \l + .05){$y_1$};
        \node[above] at (5 * \l / 30 - \l / 60 + 0.05, \l + .05){$y_2$};

        \foreach \x in {0,1,2,...,9} {
          \draw [fill=black] (\x * \l / 30 + \l / 60,\l - \x * \l / 30 - \l / 60) circle (\l / 120);
        }
        \foreach \x in {0,1,...,8} {
          \draw [fill=black] (\x * \l / 30 + \l / 30 + \l / 60,\l - \x * \l / 30 - \l / 60) circle (\l / 120);
        }
        \foreach \x in {0,1,...,8} {
          \draw [fill=black] (\x * \l / 30 + \l / 60,\l - \x * \l / 30 - \l / 60 - \l / 30) circle (\l / 120);
        }
        \draw [fill=black] (\l/2,\l - \l / 8) circle (\l / 120);
        \draw [fill=black] (\l/8,\l - \l / 2) circle (\l / 120);
        \draw [fill=black] (2*\l/3,\l - \l / 6) circle (\l / 120);
        \draw [fill=black] (\l/6,\l - 2*\l / 3) circle (\l / 120);
      \end{tikzpicture}
      \extadd 
      \begin{tikzpicture}[baseline={([yshift=-0.0ex]current bounding box.center)}]
        \def\l{1.}
        \def\o{\l/8}
        \def\h{\l/2}
        \def\q{\l/4}
        \node[below] at (\l/2,0){$\tiny C\!B_{\nu_1}$};
        \draw (0,0) rectangle (\l,\l);
        \foreach \x in {0,1,...,3} {
          \draw [fill=gray] (\x*\q,\l-\x*\q) rectangle ++(\q,-\q);
        }
        \foreach \x in {0,1} {
          \draw (\x*\h,\l-\x*\h) rectangle ++(\h,-\h);
        }

        \draw [thick, densely dotted,color=blue] (-0.1,0.6) -- (0.4,0.6);
        \draw [thick, densely dotted,color=blue] (-0.1,0.1) -- (0.4,0.1);
        \draw [thick, densely dotted,color=blue] (0.1,0.1) -- (0.1,1.1);
        \draw [thick, densely dotted,color=blue] (0.4,0.1) -- (.4,1.1);
        \draw [fill=white] (0.1,0.1) circle (\l / 50);
        \draw [fill=white] (0.1,0.6) circle (\l / 50);
        \draw [fill=white] (0.4,0.1) circle (\l / 50);
        \draw [fill=white] (0.4,0.6) circle (\l / 50);
        
        \node[left] at (-0.1,0.1){$x_2$};
        \node[left] at (-0.1,0.6){$x_1$};

        \node[above] at (0.1 - 0.05, \l + .05){$y_1$};
        \node[above] at (0.4 + 0.05, \l + .05){$y_2$};

      \end{tikzpicture}
      \extadd
      \begin{tikzpicture}[baseline={([yshift=-0.0ex]current bounding box.center)}]
        \def\l{1.4}
        \def\o{\l/8}
        \def\h{\l/2}
        \def\q{\l/4}
        \node[below] at (\l/2,0){$C\!B_{\nu_2}$};

        \draw (0,0) rectangle (\l,\l);
        \foreach \x in {0,1,...,7} {
          \draw [fill=gray] (\x*\o,\l-\x*\o) rectangle ++(\o,-\o);
        }
        \foreach \x in {0,1,...,3} {
          \draw (\x*\q,\l-\x*\q) rectangle ++(\q,-\q);
        }
        \foreach \x in {0,1} {
          \draw (\x*\h,\l-\x*\h) rectangle ++(\h,-\h);
        }
        \draw [thick, densely dotted,color=blue] (-0.1,1.15) -- (0.1,1.15);
        \draw [thick, densely dotted,color=blue] (0.1,1.15) -- (0.1,1.5);
        \draw [fill=white] (0.1,1.15) circle (\l / 60);
        \node[above] at (0.1, \l + .05){$y_2$};
        \node[left] at (-0.1, 1.15){$x_2$};

      \end{tikzpicture} \, ,
\end{equation}
		where one element $(x_1, y_2)$ corresponds to a nonzero element in the sparse matrix, and all $2 \times 2$ elements also appear in $\ylrev{C}_{{\nu_1}}$, but only one of them is part of $\ylrev{C}_{{\nu_2}}$. 
		In other words, the list $\mathcal{L}$ is converted to three separate lists, one associated with the sparse matrix, and one with each of the two child contribution blocks $\ylrev{C}_{{\nu_1}}$ and $\ylrev{C}_{{\nu_2}}$. The routine extract\_HODBF (see \cref{ssec:HODB_entry_eval}), used to extract a list of subblocks from an HOD-BF matrix, is then called twice, once as extract\_HODBF($\{ (\{x_1, x_2\}, \{y_1, y_2\}) \}, \ylrev{C}_{\nu_1}$) for the first child contribution block (with the list for this specific example), and for the second child once as extract\_HODBF($\{  (\{x_2\}, \{y_2\}) \}, \ylrev{C}_{\nu_2}$).
		Extracting the $2 \times 2$ submatrix from the HOD-BF matrix $\ylrev{C}_{{\nu_1}}$ in this case requires extracting one element $(x_1, y_1)$ from a low-rank product, one element $(x_1,y_2)$ from a dense block (leaf of the HOD-BF matrix), and extracting a $1 \times 2$ submatrix $(\{x_2\}, \{y_1,y_2\})$ from a butterfly matrix (lower left main off-diagonal block of the $\ylrev{C}_{{\nu_1}}$ HOD-BF matrix). Extraction from a butterfly matrix is explained in \cref{ssec::elem-extraction}, \cref{alg:elem_extract} and \cref{fig:extract_BF_decomp}. 
		
		
		\item Second, line~\ref{line:F11_inv} approximates $F^{-1}_{11}$ from the butterfly representation of $F_{11}$, see \cref{ssec:HODBFinversion}. 
		\item Next, lines~\ref{line:F12_construction} and~\ref{line:F21_construction}, the $F_{12}$ and $F_{21}$ front off-diagonal blocks are each approximated as a single butterfly matrix, using routines to extract elements from $A(I_{\tau}^{\text{s}},I_{\tau}^{\text{u}}) \extadd \ylrev{C}_{{\nu_1}} \extadd \ylrev{C}_{{\nu_2}}$ and $A(I_{\tau}^{\text{u}},I_{\tau}^{\text{s}}) \extadd \ylrev{C}_{{\nu_1}} \extadd \ylrev{C}_{{\nu_2}}$ respectively. For $F_{12}$, the tree $\mathcal{T}_H$ corresponding to $F_{11}$ is used as $\mathcal{T}_O$, and the tree corresponding to $F_{22}$ is used for $\mathcal{T}_S$, and vice versa for $F_{21}$. Note that we truncate the trees $\mathcal{T}_H$ if needed to enforce that $\mathcal{T}_S$ and $\mathcal{T}_O$ have the same number of levels. \Cref{ssec:separator_bisection} discusses the generation of the hierarchical partitioning. 
		\item Next, see line~\ref{line:S_construction} of \cref{alg:preconditioner}, the Schur complement update $S = F_{21} F^{-1}_{11}F_{12}$ is computed as a single butterfly matrix using randomized matrix-vector products, see \cref{ssec::random-bf-construction}. The matrix vector products can be performed efficiently, since both $F_{12}$ and $F_{21}$ are already compressed as butterfly and $F^{-1}_{11}$ is approximated as an HOD-BF matrix. 
		\item The final step for this front is to construct the contribution block of $\tau$. $\ylrev{C}_{\tau}$ as an HOD-BF matrix, again using element extraction, now from $\ylrev{C}_{{\nu_1}} \extadd \ylrev{C}_{{\nu_2}} - S$, where $\ylrev{C}_{{\nu_1}}$ and $\ylrev{C}_{{\nu_2}}$ are in HOD-BF form and $S$ is a single butterfly matrix. $S$ can be released as soon as the contribution block has been assembled, and the contribution block is kept in memory until it has been used to assemble the parent front.
	\end{enumerate}
	
	\begin{algorithm}
		\textbf{Input:} $A \in \mathbb{R}^{N \times N}$, $b \in \mathbb{R}^{N}$ \\
		\textbf{Output:} $x \approx A^{-1} b$
		\begin{algorithmic}[1]
			\State $\tilde A \leftarrow P (D_r A D_c Q_c) P^\top$  \hfill \Comment{scaling, and permutation for stability and fill reduction}
			\State \label{alg_line:sepreorder} $\hat A \leftarrow \hat{P} \tilde A \hat{P}^\top$ \hfill \Comment{rank-reducing separator reordering, \cref{ssec:separator_bisection} }
			\State build assembly tree: define $I^{\text{s}}_\tau$ and $I_\tau^{\text{u}}$ for every frontal matrix $F_\tau$
			\For{nodes $\tau$ in assembly tree in topological order}
			\If{$\text{dimension}(F_{\tau}) < n_\text{min}$}
			\State construct $F_{\tau}$ as a dense matrix \Comment{\cref{alg:mf}}
			\Else
			\State $F_{11} \leftarrow \text{HODBF\_entry\_eval}
			\left( \hat A(I_{\tau}^{\text{s}},I_{\tau}^{\text{s}})
			\extadd \ylrev{C}_{{\nu_1}} \extadd \ylrev{C}_{{\nu_2}} \right)$ \Comment{\cref{ssec:HODB_entry_eval}} \label{line:F11_HODBF}
			\State $F_{11}^{-1} \leftarrow \text{HODBF\_invert}\left(F_{11}\right)$
			\Comment{\cref{alg:hodbf_invert}} \label{line:F11_inv}
			\State $F_{12} \leftarrow \text{BF\_entry\_eval}\left(
			\hat A(I_{\tau}^{\text{s}},I_{\tau}^{\text{u}})
			\extadd \ylrev{C}_{{\nu_1}} \extadd \ylrev{C}_{{\nu_2}} \right)$ \Comment{\cref{alg:BF_entry}} \label{line:F12_construction}
			\State $F_{21} \leftarrow \text{BF\_entry\_eval}\left(
			\hat A(I_{\tau}^{\text{u}},I_{\tau}^{\text{s}})
			\extadd \ylrev{C}_{{\nu_1}} \extadd \ylrev{C}_{{\nu_2}} \right)$ \Comment{\cref{alg:BF_entry}} \label{line:F21_construction}
			\State $S \leftarrow \text{BF\_random\_matvec}
			\left(F_{21} F_{11}^{-1} F_{12}\right)$ \Comment{\cref{ssec::random-bf-construction}} \label{line:S_construction}
			\State $\ylrev{C}_{\tau} \leftarrow \text{HODBF\_entry\_eval}
			\left(\ylrev{C}_{{\nu_1}} \extadd \ylrev{C}_{{\nu_2}} - S\right)$ \Comment{\cref{ssec:HODB_entry_eval}}\label{line:CB_construction}
			\EndIf
			\EndFor
			\State $x \leftarrow \text{GMRES}(A, b, M: u \leftarrow D_c Q_c P^\top \hat{P}^\top \,\, \text{bwd-solve}(\text{fwd-solve}(\hat{P} P D_r v ) ))$ \label{line:precGMRES}
		\end{algorithmic}
		\caption{Sparse rank-structured multifrontal factorization using hierarchically off-diagonal butterfly matrix compression, followed by a GMRES iterative solve using the multifrontal factorization as an efficient preconditioner.}
		\label{alg:preconditioner}
	\end{algorithm}
	
	The final sparse rank-structured factorization can be used as an efficient preconditioner $M$ in GMRES for example, line \ref{line:precGMRES} in \cref{alg:preconditioner}. Preconditioner application requires forward and backward solve phases. The forward solve traverses the assembly tree from the leaves to the root and applies \ylrev{$F_{11}^{-1}$ followed by $F_{21}$ associated with each node $\tau$}, and then the backward solve traverses back from the root to the leafs, applying \ylrev{$F_{12}$}. Currently, we do not guarantee that the preconditioner is symmetric (or positive definite) for a symmetric (or positive definite) input matrix $A$.
	

	\subsection{Hierarchical Partitioning from Recursive Separator 
		Bisection\label{ssec:separator_bisection}}
	The butterfly partitioning, illustrated in \cref{fig:complementary_rank}, can typically be constructed by a hierarchical clustering of the source and observer point sets, $S$ and $O$, and similarly, point set coordinates can be used in clustering to define the HOD-BF partitioning hierarchy. However, in the purely algebraic setting considered here, geometry or point coordinates are not available. Instead we define the HOD-BF hierarchy of $F_{11}$ by performing a recursive bisection (not to be confused with nested dissection), using METIS, of the graph corresponding to $A(I_{\tau}^s,I_{\tau}^s)$. This defines the HOD-BF tree and a corresponding permutation of the rows/columns of $F_{11}$, and hence also the partitioning of the butterfly off-diagonal blocks of $F_{11}$. This permutation -- globally denoted as $\hat P$, see line \ref{alg_line:sepreorder} in \cref{alg:preconditioner} -- drastically reduces the ranks encountered in the off-diagonal low-rank and butterfly blocks. See \cref{fig:sep_order} for the recursive bisection, and \cref{fig:MF_BF_tree} for the corresponding HOD-BF partitioning. For the $F_{22}$ block, no such recursive bisection is performed, but the indices in $I_{\tau}^u$ are sorted and partitioned using a balanced binary tree. 

	
	\subsection{Graph Nearest Neighbor Search\label{ssec:neighbor_sampling}}
	During the graph bisection from \cref{ssec:separator_bisection}, to define the hierarchical matrix structure, edges in the graph of $A(I^s_{\tau},I^s_{\tau})$ will be cut by the partitioning. These edges correspond to nonzero entries in the off-diagonal blocks of the $F_{11}$ HOD-BF matrix. For a 2D problem, with 1D separators, there are $\mathcal{O}(1)$
	such entries, while for a 3D problem there are $\mathcal{O}(k)$
	such entries, with $k$ denoting the number of grid points along
	each dimension. As shown in \cref{eq:front_assembly}, these nonzeros are combined with the dense contribution blocks from the children fronts. However, these nonzero entries which come directly from the sparse matrix contribute significantly to the off-diagonal blocks, and to the numerical rank of these blocks. Based on the graph distance, we select for each point the $k_{\rm nn}$ nearest neighbors and pass them to the butterfly matrix construction, see \cref{ssec::extraction-bf-construction} and \cref{alg:BF_entry}. Recall that we use nearest neighbors in addition to uniform points as proxy points to accelerate the ID in \cref{alg:BF_entry}.

	More specifically, we consider the graph of $\hat A(I^s_{\tau},I^s_{\tau})$, and for each vertex in this graph we search, using a breadth-first search, for the $k_{\rm nn}$ nearest-neighbors in any of the off-diagonal blocks of the HOD-BF representation of $F_{11}$. \ylrev{This means we look at all length-$k$ connections in the graph with increasing $k$ until we pick the first $k_{\rm nn}$ vertices}. Similarly, for $F_{22}$ we look for the $k_{\rm nn}$ nearest neighbors in the graph $\hat A(I_{\tau}^u,I_{\tau}^u)$. For the main off-diagonal blocks $F_{12}$ (and $F_{21}$), we look for the nearest neighbors in the graph $\hat A(I^{s}_{\tau},I^{u}_{\tau})$ (and $\hat A(I^{u}_{\tau},I^{s}_{\tau})$) by performing a breadth-first search in the graph $\hat A(I^s_{\tau},I^s_{\tau}) \cup \hat A(I^{s}_{\tau},I^{u}_{\tau}) \cup \hat A(I_{\tau}^u,I_{\tau}^u)$. \ylrev{It is worth mentioning that generating the hierarchical partitioning and performing nearest neighbor search are computationally inexpensive.}
	
	A similar pseudo-skeleton low-rank approximation scheme based on graph distances was proposed in~\cite{aminfar2016fast}, where it is referred to as the boundary distance low-rank approximation scheme.

	\subsection{Complexity Analysis\label{ssec:cc}}
	
	
	For the complexity analysis, we consider regular $d$-dimensional meshes with $k$ gridpoints per dimension, for a total of $N = k^d$ degrees of freedom, with a stencil that is 3 points wide in each dimension. For the sparsity preserving ordering, we use nested dissection to recursively divide the mesh into $L = d\log k -\mathcal{O}(1)$ levels. At each level $\ell=0,\ldots,L$ there are $2^\ell$ separators with \ylrev{diameters (i.e., largest possible distance between two points on the separator)} of $\mathcal{O}(k/2^{\lfloor \ell/d\rfloor})$ and frontal matrices of size $\mathcal{O}(n)=\mathcal{O}((k/2^{\lfloor \ell/d\rfloor})^{d-1})$.
	For the analysis of the rank-structured solver, we split the fronts into dense and compressed fronts using a switching level $\ell_s=L-\mathcal{O}(1)$. Fronts closer to the top, i.e., at levels $\ell < \ell_s$, are typically larger and are thus compressed using the HOD-BF format, while all fronts at levels $\ell \geq \ell_s$ are stored as regular dense matrices. Note that in the implementation, we do not use a switching level, but instead decide only based on the actual size of the front. The total factorization flops $\mathcal{F}(k, d)$ and solution flops $\mathcal{S}(k, d)$ for the multifrontal solver are: \ylrev{(ignoring those at levels $\ell \geq \ell_s$ as they only scale as $\mathcal{O}(N)$)}
	\begin{align}
	\mathcal{F}(k, d) &\approx 
	\ylrev{\sum_{\ell = 0}^{\ell_s}{2^{\ell} \mathcal{F}_{\mathrm{BF}}\left( \left(\frac{k}{2^{\lfloor\ell/d\rfloor}} \right)^{d-1} \right)}}\label{eqn:F}\\
	\mathcal{S}(k, d) &= 
	\ylrev{\sum_{\ell = 0}^{\ell_s}{2^{\ell} \mathcal{S}_{\mathrm{BF}}\left( \left(\frac{k}{2^{\lfloor\ell/d\rfloor}} \right)^{d-1} \right)}}\label{eqn:S}.
	\end{align}
	Here $\mathcal{F}_{\mathrm{BF}}(n)$ and $\mathcal{S}_{\mathrm{BF}}(n)$ (\ylrev{$\leq\mathcal{F}_{\mathrm{BF}}(n)$}) denote the cost of factorization (including construction) and solution of a HOD-BF compressed front of size $\mathcal{O}(n)$. In addition, it is straightforward to verify that the memory requirement of the multifrontal solver $\mathcal{M}(k, d)\sim\mathcal{S}(k, d)$ as the solution phase typically requires a single-pass of the memory storage. Recall that the exact multifrontal solver has $\mathcal{F}=\mathcal{O}(N^2)$, $\mathcal{S}=\mathcal{O}(N^{4/3})$ for $d=3$ and $\mathcal{F}=\mathcal{O}(N^{3/2})$, $\mathcal{S}=\mathcal{O}(N\log N)$ for $d=2$. \ylrev{As we will see next, lower complexities can be achieved as long as $\mathcal{F}_{\mathrm{BF}}(n)<\mathcal{O}(n^3)$ (see Table 2.1 of \cite{Theo19bridging})}. 
	
	In what follows, we derive the complexity of the HOD-BF multifrontal solver and compare with the HSS multifrontal solver in~\cite{xia2013randomized} for both high-frequency and low-frequency wave equations. Here ``high-frequency" refers to linear systems whose size is proportional to certain power of the wavenumber (e.g., by fixing the number of grid points per wavelength to $\mathcal{O}(10)$), while ``low-frequency" refers to linear systems whose size is, roughly speaking, independent of the wavenumber. We choose the high-frequency Helmholtz equation and the Poisson equation, both in homogeneous media, as two representative cases. Note that the proposed solver can be applied to a much wider range of wave equations and media with low complexities. Let $r(n)$ denote the maximum rank of the HOD-BF or HSS representation of a front of size $\mathcal{O}(n)$. As the front represents a numerical Green's function that resembles the free-space Green's function of the wave equations, we claim without proof that the rank $r(n)$ also behaves similarly to that arising from boundary
	element methods~\cite{Yang_2020_BFprecondition,liu2020butterfly}.
	For more rigorous proofs regarding ranks in the frontal matrices,
	see~\cite{Engquist2018Greenfunction}. We further assume (and observed) that the rank in HOD-BF or HSS representation of the front remains similar after the inversion process. 
	
	\begin{table*} 
		\footnotesize
		\centering	
		\begin{tabular}{|c|c|cc|cc|cc|c|}
			\hline		
			& & \multicolumn{2}{c}{rank $r(n)$}  & \multicolumn{2}{|c}{factor flops $\mathcal{F}$} & \multicolumn{2}{|c|}{solve flops $\mathcal{S}$} \\
			problem & dim & HOD-BF & HSS & HOD-BF & HSS & HOD-BF & HSS \\
			\hline			\hline
			\multirow{2}{*}{Helmholtz} & $2$ & $\log n$ & $n$ & $N$& $N^{3/2}$& $N$ & $N\log N$  \\
			\cline{3-8}		
			\cline{2-6}				
			& $3$ & $n^{1/4}$ & $n$ &  $N\log^2 N$ &  $N^{2}$ & $N$  & $N^{4/3}$ \\
			\hline		
			\hline		
			\multirow{2}{*}{Poisson}& $2$ & $\log n$ & $\log n$  &  $N$ &  $N$ & $N$  & $N$ \\
			\cline{2-8}		
			& $3$ & $n^{1/4}$ & $n^{1/2}$ &  $N\log^2 N$ &  $N^{4/3}$ & $N$ & $N$ \\

			\hline
		\end{tabular}
		\vspace{-10pt}
		\caption{Asymptotic complexity of the HOD-BF and HSS multifrontal solvers for 2D and 3D, Helmholtz and Poisson equations. The $\mathcal{O(\cdot)}$ has been dropped. Here $n$ denotes the size of a front and $N$ is the global number of degrees of freedom in the sparse system.}\label{tab:cc}
		\vspace{-25pt}
	\end{table*}

	\paragraph{Helmholtz equation} Consider the $F_{12}$ and $F_{21}$ blocks of a front $F$ of size $\mathcal{O}(n)$ which represent the numerical Green's function interaction between two crossing separators. See \cref{fig::mesh_11} 
	for an illustration of such an interaction between two crossing separators, for instance $S_0^1$ and $S^0$. By direct application of the results in Section 3.3.2 in \cite{Yang_2020_BFprecondition} and Section 4.5.2 in \cite{liu2020butterfly} for 2D and 3D free-space Green's functions, one can show that $r(n)=\mathcal{O}(\log n)$ for $d=2$ and $r(n)=\mathcal{O}(n^{1/4})$ for $d=3$. \ylrev{Irrespective of whether $d=2$ or $d=3$}, the costs of construction from entry evaluation and randomized matvec still scale \ylrev{respectively} as $\mathcal{O}(n\log^2 n)$ and $\mathcal{O}(n^{3/2}\log n)$ just like the constant rank case in \cref{ssec::elem-extraction} and \cref{ssec::random-bf-construction}. \\
	\ylrev{\textbf{Remark.} The rank of $F_{12}$ and $F_{21}$ representing interactions between crossing separators in 3D may grow faster than $r(n)=\mathcal{O}(n^{1/4})$ when the frequency is high enough. As a remedy, one can either modify the nested dissection algorithm (for regular domains) to generate parallel separators for the top few levels of the assembly tree, or consider only well separated submatrices of $F_{12}$/$F_{21}$ as butterflies. This assures constant rank for $F_{12}$/$F_{21}$ and $\mathcal{O}(n^{1/4})$ rank for $F_{11}$. }     
	
	We summarize the computational complexities of lines \ref{line:F11_HODBF} to \ref{line:CB_construction} of \cref{alg:preconditioner} here: BF\_entry\_eval at \ref{line:F12_construction} and \ref{line:F21_construction} requires $\mathcal{O}(n\log^2 n)$ operations, HODBF\_entry\_eval at \ref{line:F11_HODBF} and \ref{line:CB_construction} requires $\mathcal{O}(n\log^3 n)$ operations, BF\_random\_matvec at line \ref{line:S_construction} requires $\mathcal{O}(n^{3/2}\log n)$ operations, and HODBF\_invert at line \ref{line:F11_inv} requires $\mathcal{O}(n^{3/2}\log n)$ operations. In addition, the corresponding storage cost requires $\mathcal{O}(n\log^2 n)$ memory units. Therefore, the cost of factorization and solution of a HOD-BF compressed front is $\mathcal{F}_{\mathrm{BF}}(n)=\mathcal{O}(n^{3/2}\log n)$ and $\mathcal{S}_{\mathrm{BF}}(n)=\mathcal{O}(n\log^2 n)$. Plugging these estimates into \cref{eqn:F} and \cref{eqn:S} will yield the total factorization and solution cost of the HOD-BF multifrontal solver as
	
	\begin{align}
	\mathcal{F}(k, 2) &\ylrev{\approx
		\sum_{\ell = 0}^{\ell_s}{k^{2}\frac{2^{\lfloor\ell/4\rfloor}}{k^{1/2}}\log\left(\frac{k}{2^{\lfloor\ell/2\rfloor}}\right)} \overset{\frac{k}{2^{\lfloor\ell/2\rfloor}}\rightarrow 2^t}{=\joinrel=}\sum_{t=0}^{\log (k)}k^2\frac{t}{2^{t/2}}=k^2=N}\label{eqn:cc_bf_factor2}\\
	\mathcal{F}(k, 3) &\ylrev{\approx
		\sum_{\ell = 0}^{\ell_s}k^{3}  \log\left( \frac{k}{2^{\lfloor\ell/3\rfloor}}\right) =k^3\log^2 k= \frac{1}{9} N\log^2N}\label{eqn:cc_bf_factor3}\\
	\mathcal{S}(k, d) &\ylrev{\approx
		\sum_{\ell = 0}^{\ell_s}{k^{d}\frac{2^{\lfloor\ell/d\rfloor}}{k}\log^2\left(\frac{k}{2^{\lfloor\ell/d\rfloor}}\right)} \overset{\frac{k}{2^{\lfloor\ell/d\rfloor}}\rightarrow 2^t}{=\joinrel=} \sum_{t=0}^{\log (k)}k^d\frac{t^2}{2^{t}}=k^d=N}.\label{eqn:cc_bf_solve}
	\end{align}
	Note that $\mathcal{O(\cdot)}$ has been dropped in the above equations. Hence, the HOD-BF multifrontal solver can attain quasi-linear complexity for high-frequency Helmholtz equations. In contrast, one can show, based on the arguments in \cite{Bucci1987SpatialBandwidth, Engquist2018Greenfunction}, that the HSS rank $r(n)=\mathcal{O}(n)$ for both $d=2$ and $d=3$ due to the highly-oscillatory interaction between two crossing separators, which yields   
	$\mathcal{O}(n^3)$ and $\mathcal{O}(n^2)$ factorization and solution complexity for one front and hence no asymptotic gains using the HSS multifrontal solver compared to exact multifrontal solvers. We summarize these complexities in \cref{tab:cc}.
	
	\paragraph{Poisson equation} The complexity of the HOD-BF multifrontal solver for the Poisson equation can be estimated similarly to the Helmholtz equation. First, one can show that the butterfly rank $r(n)=\mathcal{O}(\log n)$ for $d=2$ and $r(n)=\mathcal{O}(n^{1/4})$ for $d=3$, just like the Helmholtz case. This yields similar complexities as those in \cref{eqn:cc_bf_factor2,eqn:cc_bf_factor3,eqn:cc_bf_solve} with smaller leading constants. For comparison, the HSS rank behaves as $r(n)=\mathcal{O}(\log n)$ for $d=2$ and $r(n)=\mathcal{O}(n^{1/2})$ for $d=3$ (see \cite{Ho2012rs,chandrasekaran2010numerical,Engquist2018Greenfunction}), which yields fast HSS mutifrontal solvers. We refer the readers to \cite{xia2013randomized} for detailed analysis and list the complexities in \cref{tab:cc}. One can see that lower complexity can be attained using HOD-BF multifrontal ($\mathcal{O}(N\log^2 N)$) than HSS multifrontal ($\mathcal{O}(N^{4/3})$) for the factorization when $d=3$ (\ylrev{Note that multifrontal-like solvers with other compression formats such as HIF \cite{ho2016hierarchical} can also attain quasi-linear complexities}); similar complexities are attained for all the other entries in the table, despite that HOD-BF multifrontal can yield larger leading constants than HSS multifrontal.

	\section{Experimental Results\label{sec:example}}
	Experiments reported here are all performed on the Haswell nodes of the Cori machine, a Cray XC40, at NERSC in Berkeley. Each of the $2,388$ Haswell nodes has two $16$-core Intel Xeon E5-2698v3 processors and 128GB of 2133MHz DDR4 memory. We developed a distributed memory code but we omit the description of the parallel algorithms here and will discuss this in a future paper. 

	The approximate multifrontal solver is used as a preconditioner for restarted GMRES($30$) with modified Gram-Schmidt and a zero initial guess. All experiments are performed in double precision with absolute or relative stopping criteria $\| u_i\| \leq 10^{-10}$ or $\| u_i\| / \|u_0\| \leq 10^{-6}$, where $u_i = M^{-1}(A x_i - b)$ is the preconditioned residual, with $M$ the approximate multifrontal factorization of $A$. For the exact multifrontal solver, we use iterative refinement instead of GMRES. For the tests in \cref{ssec::visco-elastic,ssec::P3}, the nested dissection ordering is constructed from planar separators. For the test in \cref{ssec::indefinite-Maxwell} the nested dissection ordering from METIS~\cite{karypis1998fast} was used. For all the tests the column permutation and row/column scaling were disabled.
	
	\ylrev{For each problem, we compare three types of multifrontal solvers:
		``Exact''--no compression, ``HSS($\varepsilon$)''--HSS compression
		with tolerance $\varepsilon$, and ``HOD-BF($\varepsilon$)''--HOD-BF with
		tolerance $\varepsilon$.}

	\subsection{Visco-Acoustic Wave Propagation\label{ssec::visco-elastic}}
	We first consider the 3D visco-acoustic wave propagation governed by the Helmholtz equation
	\begin{equation}
	\left(\sum_i\rho({\bf x})\frac{\partial}{\partial x_i}\frac{1}{\rho({\bf x})}\frac{\partial}{\partial x_i}\right)p({\bf x})+\frac{\omega^2}{\kappa^2({\bf x})}p({\bf x})=-f({\bf x}).\label{eq:visco-acoustic}
	\end{equation}
	Here ${\bf x}=(x_1,x_2,x_3)$, $\rho(\bf x)$ is the mass density, $f({\bf x})$ is the acoustic excitation, $p(\bf x)$ is the pressure wave field, $\omega$ is the angular frequency, $\kappa({\bf x})=v({\bf x})(1-{i}/(2q({\bf x})))$ is the complex bulk modulus with the velocity $v({\bf x})$ and quality factor $q({\bf x})$. We solve \cref{eq:visco-acoustic} by a finite-difference discretization on staggered grids using a 27-point stencil and 8 PML absorbing boundary layers~\cite{operto20073d}. This requires direct solution of a sparse linear system where each matrix row contains 27 nonzeros, whose values depend on the coefficients and frequency in \cref{eq:visco-acoustic}.   
	
	\begin{table}[h!]
		\footnotesize
		\begin{center}
			\begin{tabular}{|r||c||c|c||c|c|}
				\hline
				Solver & Exact & HSS & HOD-BF & HOD-BF & HOD-BF \\
				$\varepsilon$ & - & $10^{-3}$ & $10^{-3}$ & $10^{-2}$ &  $10^{-3}$\\				
				$n_{\rm min}$ & - & 10K & 10K & 10K & 7K \\				
				\hline
				Compressed fronts & 0 & 39 & 39 & 39 & 197 \\
				Dense fronts & 1,869,841 & 1,869,802 & 1,869,802 & 1,869,802 & 1,869,644\\
				Factor time (sec) & 513 & 947 & 433 & 354 &  556\\
				Factor flops ($10^{15}$) & 13.4  & 4.98  & 2.44  & 2.24  & 1.21 \\
				Flop Compression (\%) & 100 & 37.1 & 18.2 & 16.7 & 9.0\\
				Factor mem ($10^3$ GB) & 1.48  & 0.84  & 0.73  & 0.72  & 0.47 \\
				Mem Compression (\%) & 100 & 56.8 & 49.6 & 48.8 & 32.2 \\
				Max. rank & - & 4698 & 364 & 153 & 389 \\
				\hline
				Top 2 fronts & & & & & \\ 
				Mem Compression (\%) & - & 21.9/14.6 & 7.29/3.54 & 4.4/1.89 & 6.3/3.6 \\
				Rank  & - & 4538/4698 & 154/242 & 121/213 & 177/255 \\
				Front time (sec) & 37/108 & 172/195 & 52/88 & 42/60 &  46/70.6\\
				\hline
				GMRES its. & 1 & 18 & 6 & 56 & 23 \\
				Solve flops ($10^{12}$)  & 0.46  & 8.01  & 2.84  & 23.4  & 7.18 \\
				Solve time (sec) & 0.72 & 19.2 & 3.5 & 30.1  & 13.2  \\
				\hline
			\end{tabular}
		\end{center}
		\caption{Results for applying HOD-BF, \ylrev{HSS} and exact
			multifrontal solvers to \cref{eq:visco-acoustic} with constant
			coefficients and $N=250^3$. \ylrev{Here, the fronts with dimensions
				smaller than $n_{\rm min}$ are treated as dense.} For the top 2 fronts,
			we give the compression rate, maximum rank and time spent,
			separated by "/". \ylrev{The flop (or memory) compression rate is
				ratio, in percentage, of the factor flops (or memory) in compressed
				formats over that of the exact form}.
			We use $32$ compute nodes, with \ylrev{$8$} MPI ranks per node and
			\ylrev{$4$} OpenMP threads per MPI process.
		}
		\label{tab:HH}
		\vspace{-10pt}
	\end{table}
	
\begin{figure}[h!]
		\centering
		\begin{subfigure}[t]{.32\textwidth}
			\centering
			\includegraphics[width=\linewidth]{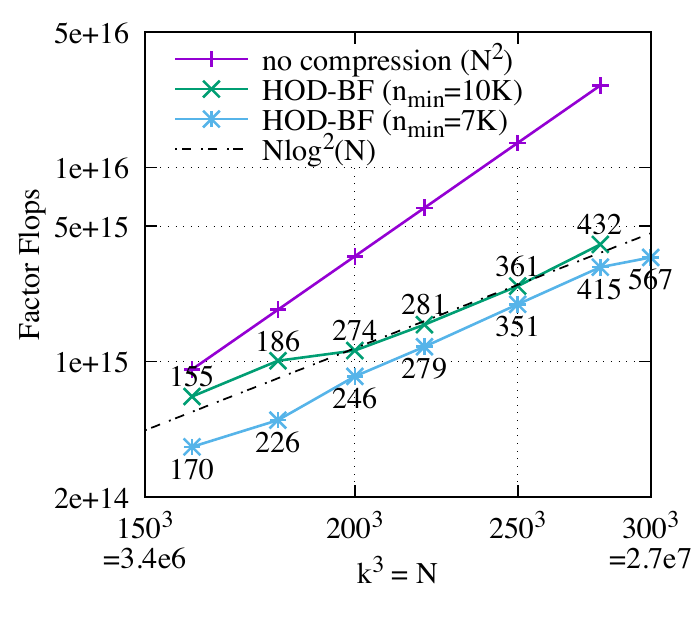}
			\vspace{-50pt}
			\caption{\label{fig:visco_acoustic_factor_flops}}
		\end{subfigure}
		\begin{subfigure}[t]{.32\textwidth}
			\centering
			\includegraphics[width=\linewidth]{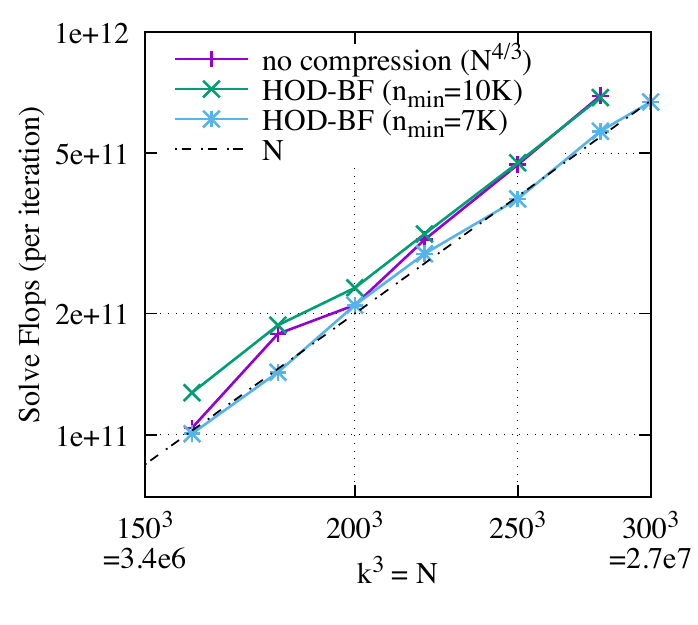}
			\vspace{-50pt}			
			\caption{\label{fig:visco_acoustic_solve_flops}}
		\end{subfigure}
		\begin{subfigure}[t]{.32\textwidth}
			\centering
			\includegraphics[width=\linewidth]{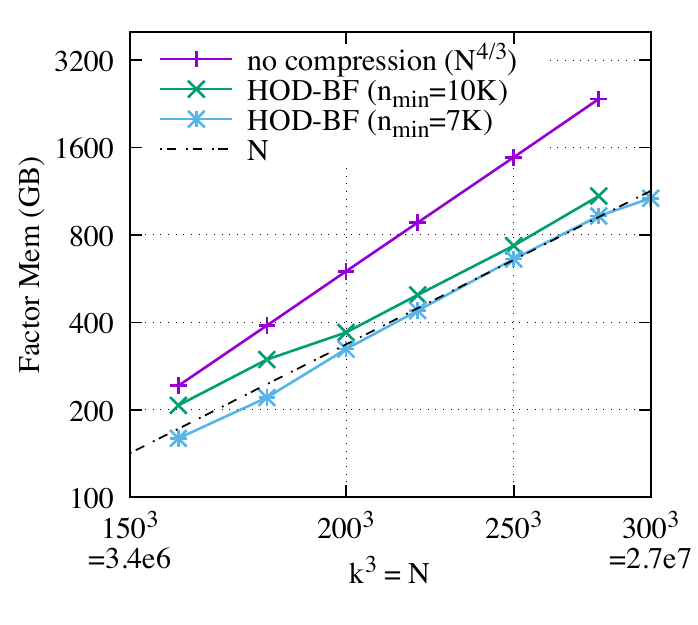}
			\vspace{-50pt}			
			\caption{\label{fig:visco_acoustic_memory}}
		\end{subfigure}
		\begin{subfigure}[t]{.32\textwidth}
			\centering
			\includegraphics[width=\linewidth]{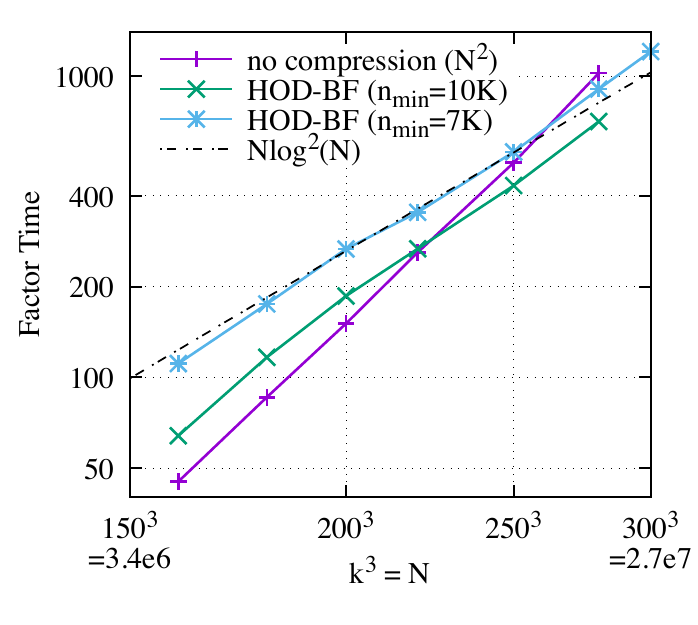}
			\vspace{-50pt}
			\caption{\label{fig:visco_acoustic_factor_time}}
		\end{subfigure}
		\begin{subfigure}[t]{.32\textwidth}
			\centering
			\includegraphics[width=\linewidth]{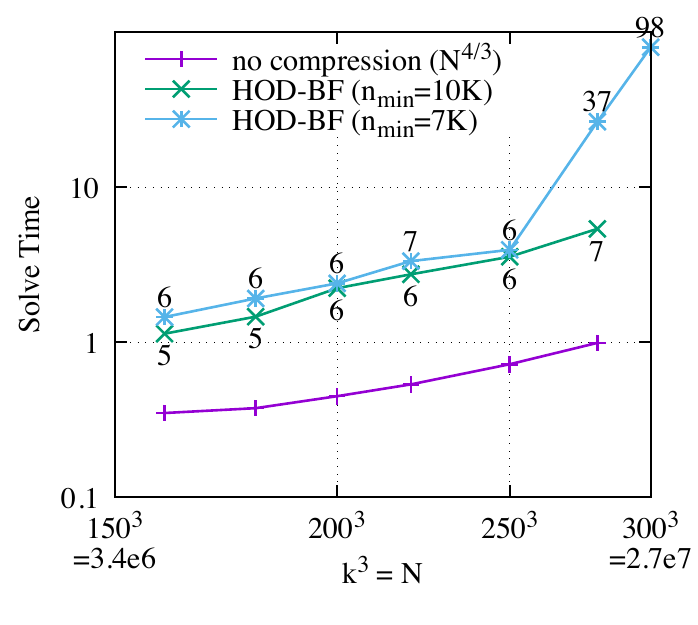}
			\vspace{-50pt}			
			\caption{\label{fig:visco_acoustic_solve_time}}
		\end{subfigure}
		\begin{subfigure}[t]{.32\textwidth}
			\centering
			\includegraphics[width=\linewidth]{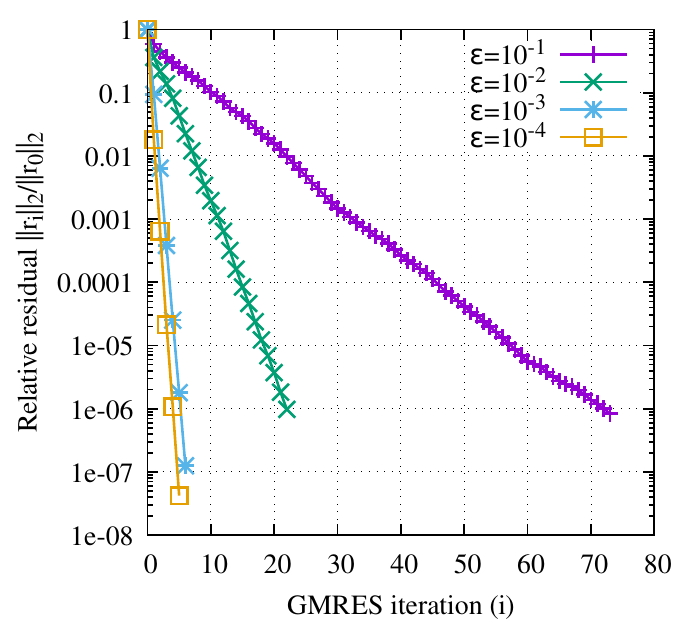}
			\vspace{-50pt}			
			\caption{\label{fig:visco_acoustic_gmres}}
		\end{subfigure}	
		\vspace{-5pt}
		\caption{\ylrev{Results for high frequency 3D Helmholtz using the exact solver with iterative refinement and the HOD-BF($10^{-3}$) multifrontal solver with GMRES. (a) Flop counts for factorization. The maximum ranks are shown at every datapoint of HOD-BF. (b) Flop counts for solve per iteration in GMRES. (c) Memory usage for the factors (not the peak working memory). (d) CPU time for factorization. (e) CPU time for solve. The number of GMRES iterations are shown at every datapoint of HOD-BF. (f) GMRES convergence for $k=200$, with different compression tolerances $\varepsilon$.} \label{fig:visco_acoustic}} 
	\end{figure}

	\paragraph{Homogeneous media} 
	We consider a cubed domain with $v({\bf x})=4000$m/s, $\rho({\bf x})=1$kg/m$^3$, $q({\bf x})=10^4$. The frequency is set to $\omega=8\pi$Hz and the grid spacing is set such that there are $15$ grid points per wavelength. First, we consider a problem with size $N=k^3$, $k=250$ and compare the performance of the HOD-BF multifrontal solver with the exact multifrontal solver \ylrev{and HSS multifrontal solver} by setting tolerances $\varepsilon = \ylrev{10^{-2}, 10^{-3}}$ and varying switching levels (corresponding to minimum compressed separators with sizes $n_{\rm min}=$ \ylrev{10K, 7K}). Here, $\varepsilon$ refers to the ID tolerance used in BF\_entry\_eval and BF\_random\_matvec. \cref{tab:HH} lists the time, flop counts, memory and ranks for the factor and solve phases, as well as those for the top two fronts.
	\ylrev{Comparing the first three columns, HOD-BF requires significantly less factor time, flops and memory than exact and HSS solvers, as well as much smaller ranks compared to HSS solvers. This is particularly the case for the top level fronts. It's also worth-mentioning that varying $\varepsilon$ in HOD-BF (column 3 and 4) leads to a trade-off between factor time and
		solve time; varying $n_{\rm min}$ in HOD-BF (column 3 and 5) leads to
		a trade-off between factor time and factor memory.}
	Next, we validate the complexity estimates in \cref{tab:cc} when varying $N$ from \ylrev{$160^3$ to $300^3$ (and correspondingly domain size from 10.7 to 20 in wavelength)}, while compressing all fronts corresponding to separators larger than \ylrev{10K}. Compared to the $\mathcal{O}(N^2)$ computation and $\mathcal{O}(N^{4/3})$ memory complexities using the exact multifrontal solver, we observe the predicted $\mathcal{O}(N\log^2N)$ computation and $\mathcal{O}(N)$ memory complexities using the HOD-BF multifrontal solver with $\varepsilon=10^{-3}$ \ylrev{(see \cref{fig:visco_acoustic_factor_flops}-\cref{fig:visco_acoustic_solve_time}). The maximum ranks and iteration counts are also shown in \cref{fig:visco_acoustic_factor_flops} and \cref{fig:visco_acoustic_solve_time}, respectively. Note that HOD-BF outperforms exact solver for $k>220$ and all data points in terms of factor time and factor memory, respectively}. Finally, we investigate the effect of HOD-BF compression tolerance on the GMRES convergence using $N=200^3$. The GMRES residual history with different $\varepsilon$ are plotted in \cref{fig:visco_acoustic_gmres}.

\begin{figure}
		\centering
		\vspace{-7.5pt}
		\begin{subfigure}[t]{.4\textwidth}
			\centering
			\includegraphics[width=\linewidth]{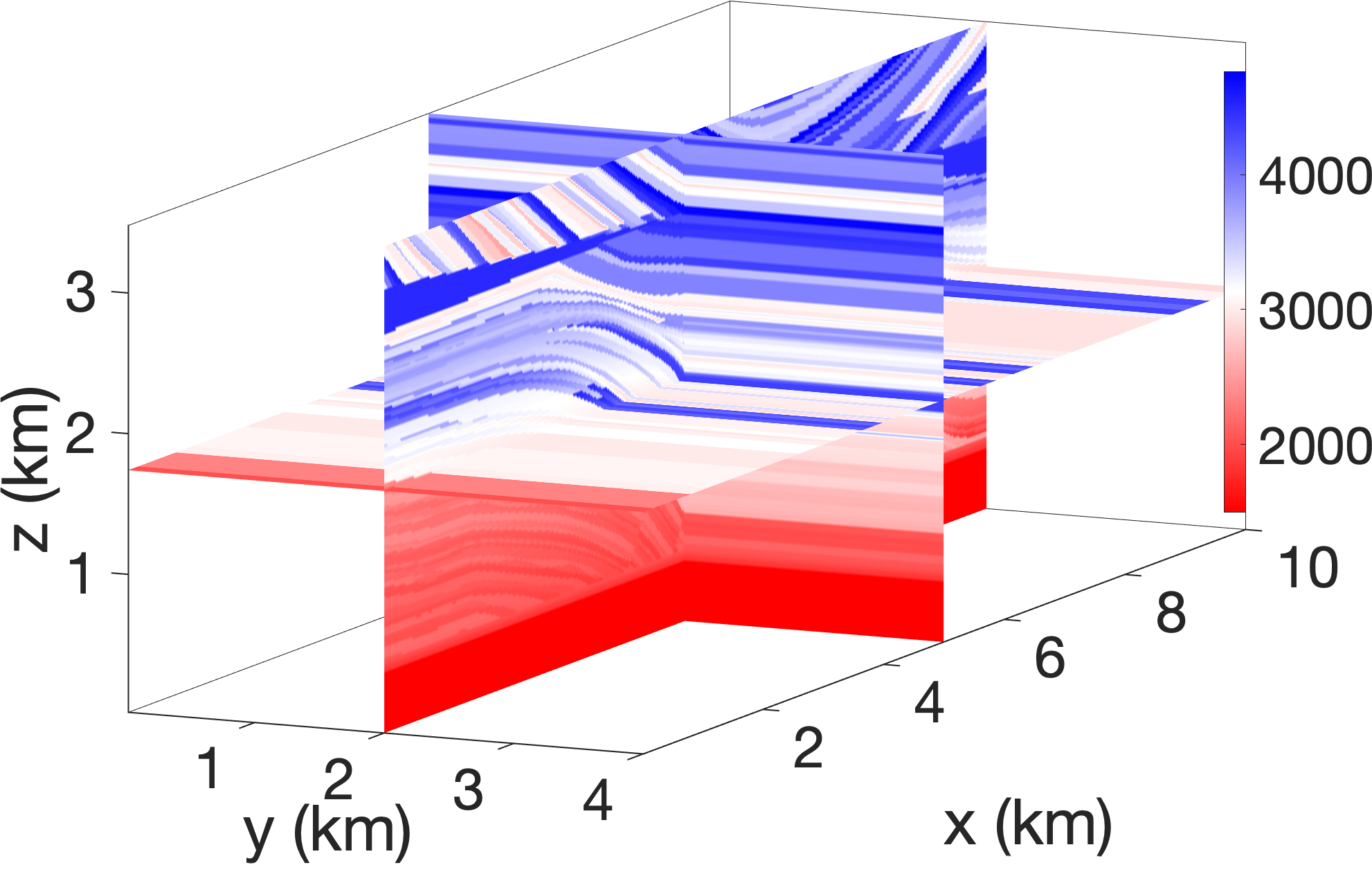}
		\end{subfigure}
		\begin{subfigure}[t]{.4\textwidth}
			\centering
			\includegraphics[width=\linewidth]{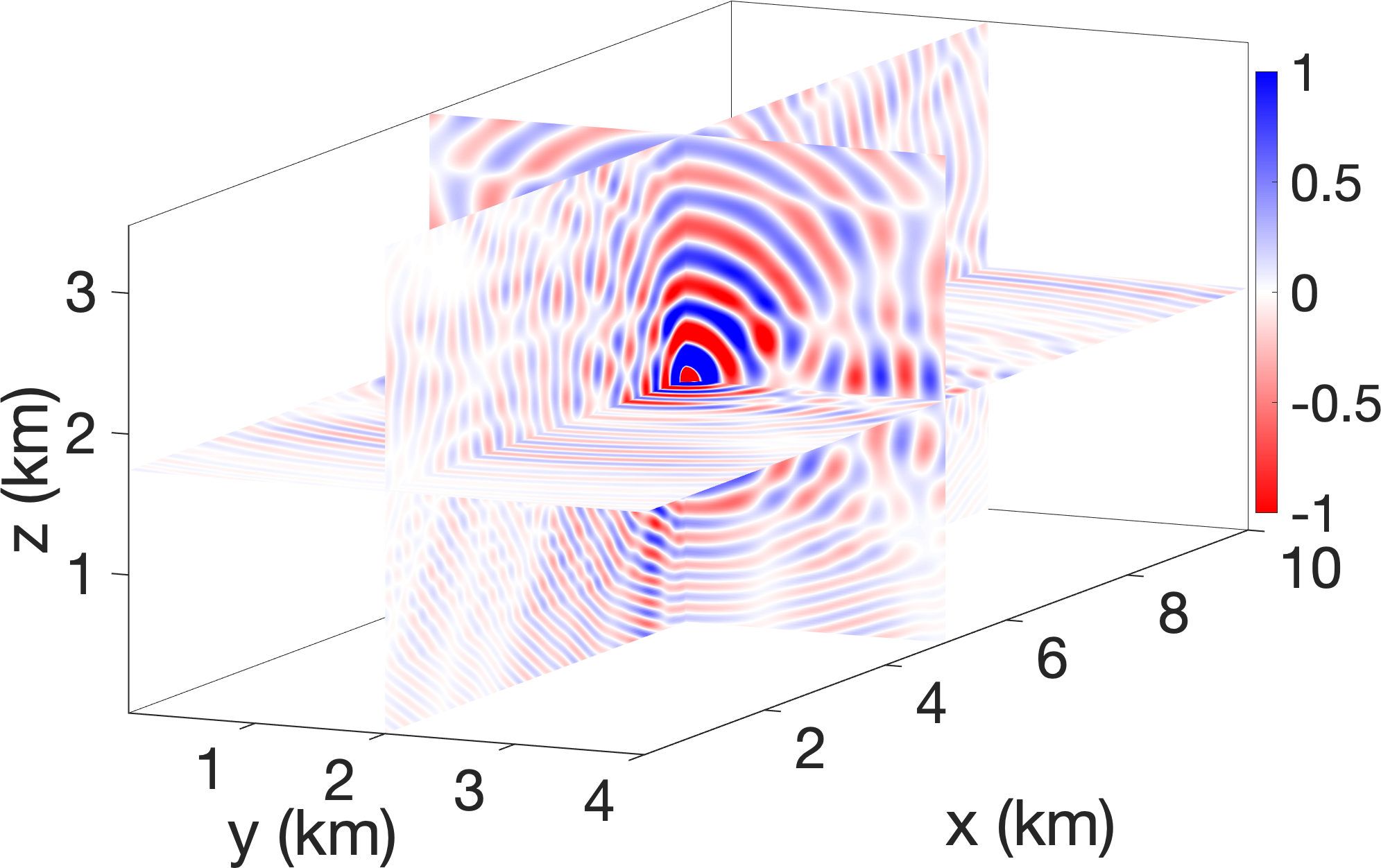}
		\end{subfigure}	
		\begin{subfigure}[t]{.4\textwidth}
			\centering
			\includegraphics[width=\linewidth]{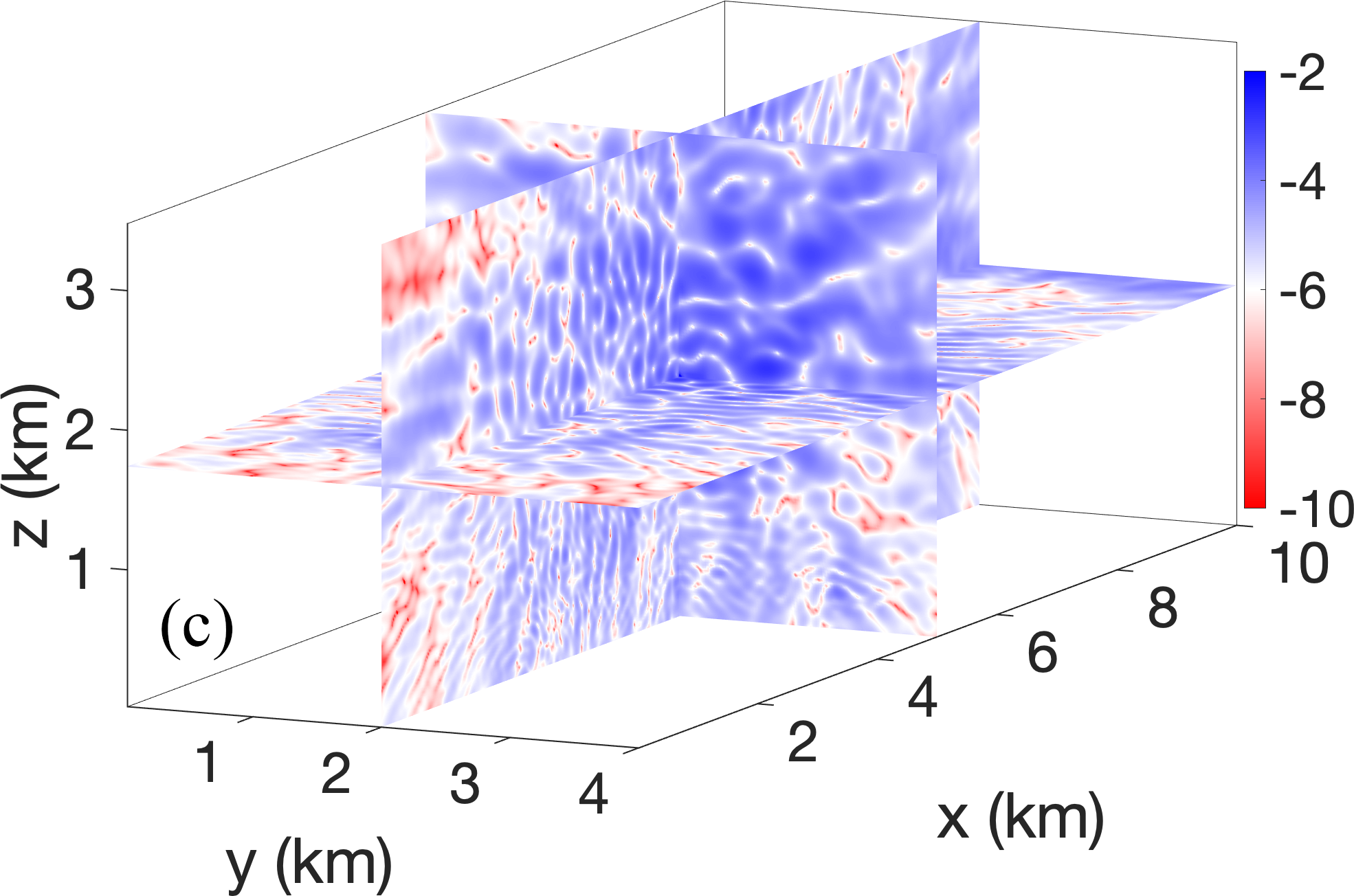}
		\end{subfigure}	
		\begin{subfigure}[t]{.4\textwidth}
			\centering
			\includegraphics[width=\linewidth]{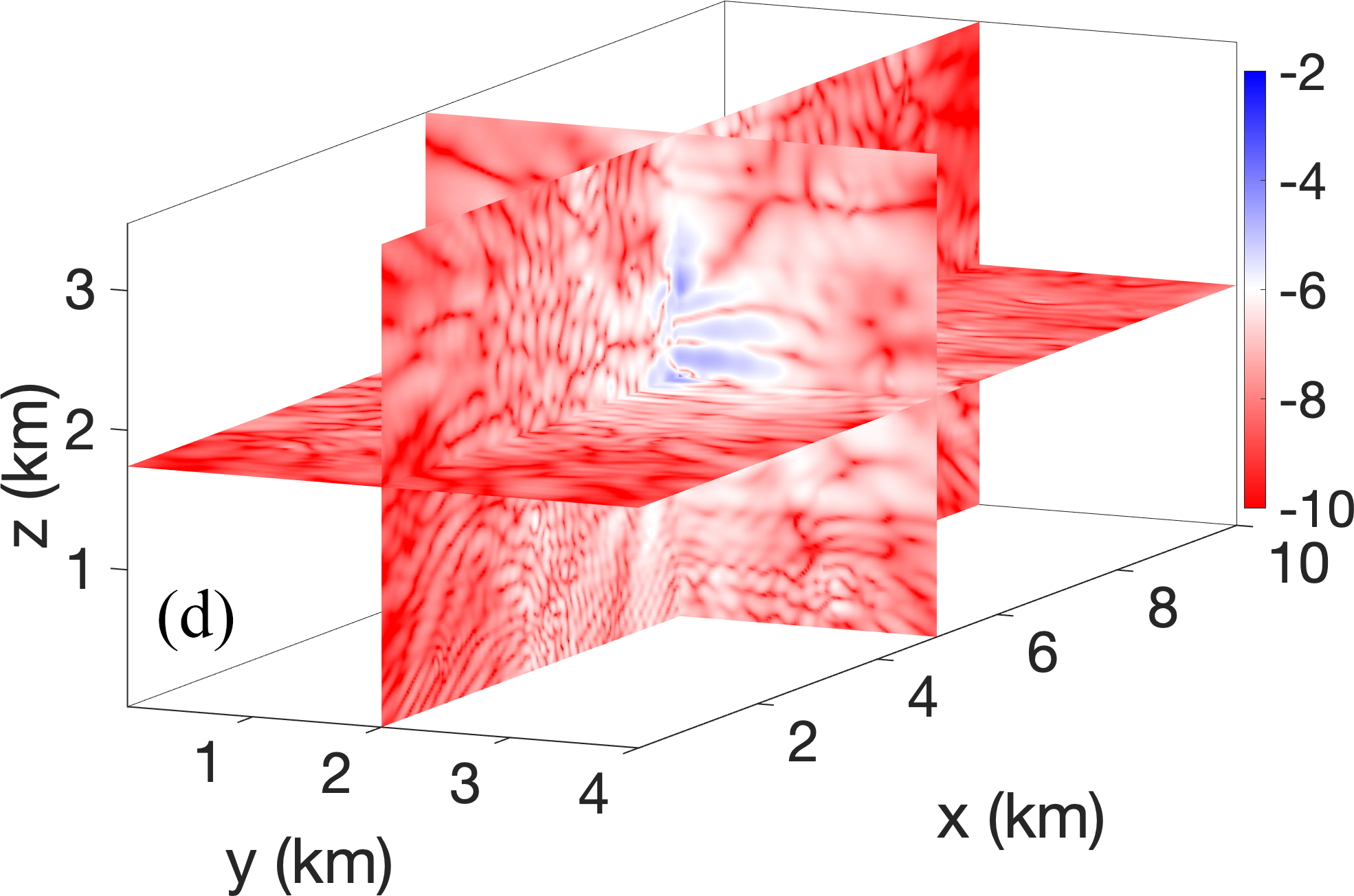}
		\end{subfigure}	
		\vspace{-5pt}
		\caption{\ylrev{(a) 3D extension of the Marmousi2 velocity model, (b) the real part of the pressure wave field $p({\bf r})$ excited by a point source at the domain center computed by the HOD-BF($10^{-4}$) multifrontal solver with GMRES,
				(c) difference in $|p({\bf r})|$ of log scale computed by
				HOD-BF($10^{-4}$) with and without GMRES, (d) difference in
				$|p({\bf r})|$ of log scale computed by HOD-BF($10^{-6}$)
				with and without GMRES.} We use 32 compute nodes.
			\label{fig:visco_acoustic_model}}
		\vspace{-20pt}
	\end{figure}

	\paragraph{Heterogeneous media} Here we use the Marmousi2 \cite{Martin2006Marmousi2} P-wave velocity model for $v({\bf x})$, and set $\rho({\bf x})=1$kg/m$^3$, $q({\bf x})=10^4$. We generate a $174\times500$ grid in the x-z plane using the Marmousi2 model and duplicate the model 200 times in the y direction, yielding a mesh of $190\times216\times516$ with $N=21,\!176,\!640$ and grid spacing 20m including the PMLs (see \cref{fig:visco_acoustic_model}). We set the frequency to $\omega=20\pi$ corresponding to 7.5 grid points per miminum wavelength. The real part of the pressure field, induced by a point source located at the domain center, is computed by the proposed HOD-BF multifrontal solver with 32 compute nodes and plotted in \cref{fig:visco_acoustic_model}. \ylrev{The difference between the pressure field, computed by the HOD-BF multifrontal solver with and without GMRES is also plotted. When $\varepsilon = 10^{-6}$, the HOD-BF multifrontal solver can also serve as a good direct solver (without GMRES).} The technical data with different compression tolerances and switching levels is listed in \cref{tab:visco_acoustic_model} \ylrev{and compared with exact and HSS multifrontal solvers}. Significant \ylrev{flop and} memory compression ratios have been observed. Note that there is a trade-off between the factor and solve times when using different tolerances and switching levels.

	\begin{table}[h!]
		\footnotesize
		\begin{center}
			\setlength\tabcolsep{1.5pt}
			\begin{tabular}{|r||c|c|c|c|c|c|c|}\hline
				Solver  & Exact & HSS & HOD-BF & HOD-BF & HOD-BF  & HOD-BF  \\	
				$\varepsilon$  & - &$10^{-3}$ & $10^{-3}$ & $10^{-3}$ & $10^{-4}$ & $10^{-6}$\\
				$n_{\rm min}$  & - & 75K &38.5K & 75K & 75K  & 75K  \\
				\hline			
				Compressed fronts & - & 143 &435 & 143 &  143 &  143 \\
				Dense fronts & 2,102,917 & 2,102,774 & 2,102,482 & 2,102,774 &  2,102,774  &  2,102,774 \\
				Factor time (sec) & 660 & 1575 & 1037 & 674 &   1049  &   1657 \\
				Factor flops ($10^{15}$)  & 17.8  & 7.33   & 2.19  & 2.17  & 2.71  & 4.51 \\
				Flop Compression (\%) & 100 & 41.2 & 12.3 & 12.2 & 15.2 & 25.3\\				
				Memory ($10^3$ GB)        & 1.97  & 1.01   & 0.58  & 0.77  & 0.8   & 0.87 \\
				Mem Compression (\%) & 100 & 51.08 & 29.91 & 39.3 & 40.7 & 44.6\\
				Maximum rank  &- & 4105 & 549 & 492 &  608 & 824 \\
				GMRES iterations & 0& 6 & 59 & 63 &  12 & 3\\
				Solve flops ($10^{12}$)   & 0.55  & 3.91   & 23.3  & 34.6  & 6.97  & 2.30 \\
				Solve time (sec) & 0.9 & 11.8 &48.7 & 52.8 &  11.0   &  4.1\\
				\hline
			\end{tabular}				
		\end{center}
		\vspace{-5pt}
		\caption{\ylrev{Data for applying HOD-BF, \ylrev{HSS and exact multifrontal}
				solvers to \cref{eq:visco-acoustic} with the Marmousi2 velocity model.}
			We use 32 compute nodes.
			\label{tab:visco_acoustic_model}}
		\vspace{-20pt}	
	\end{table}

	\subsection{Indefinite Maxwell\label{ssec::indefinite-Maxwell}}
	
\begin{figure}
		\begin{subfigure}[b!]{.54\textwidth}
			\centering
			{ \footnotesize
				\begin{tabular}{|r||c|c|}
					\hline
					Solver & \pgrev{Exact} & HOD-BF \\
					$\varepsilon$ & - & $10^{-5}$ \\
					$n_{min}$ & - & 15K \\
					\hline			
					Compressed fronts & 0 & 6  \\
					Dense fronts & 3,773,221 & 3,773,215  \\
					Factor time (sec) & 301.34 & 379.50 \\
					Factor flops ($10^{15}$)  & 2.13  &  1.28   \\
					Flop Compr. (\%) & 100 & 60.1 \\
					Memory (GB) & 541 & 426 \\
					Mem Compr. (\%) & 100 & 78.8 \\
					Max rank/front size & - & 955 / 78203  \\
					GMRES its. & 1 & 21 \\
					Solve flops ($10^{12}$) & 1.18   & 2.45  \\
					Solve time (sec) & 1.09 & 18.98 \\
					\hline
			\end{tabular}}
		\end{subfigure}
		\begin{subfigure}[b!]{.45\textwidth}
			\centering
			\includegraphics[width=.8\linewidth]{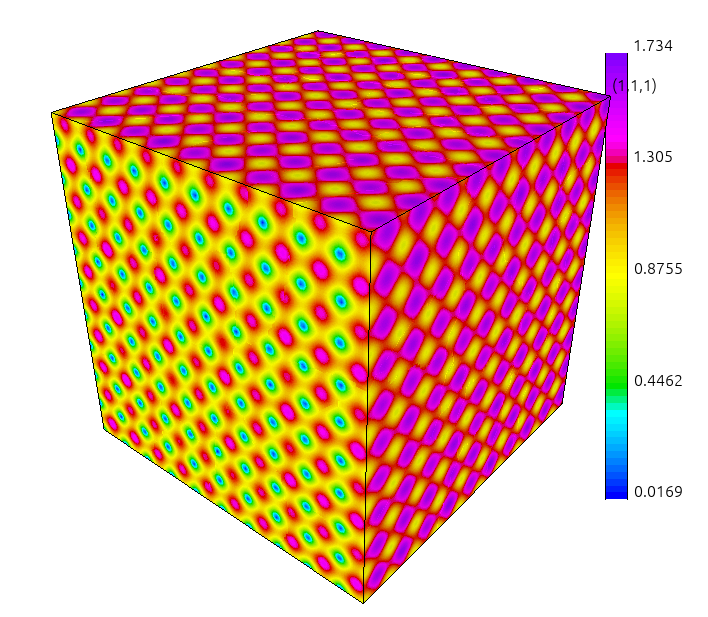}
		\end{subfigure}
		\vspace{-5pt}
		\caption{Left: Data for applying the \pgrev{exact and} HOD-BF($10^{-5}$)
			solvers to the indefinite Maxwell equation. Right: Magnitude of computed solution $E$. \pgrev{We use $16$ compute nodes.}
			\label{fig:Maxwell}}
		\vspace{-30pt}
	\end{figure}

	We solve the electromagnetics problem corresponding to the second order
	Maxwell equation, $\nabla \times \nabla \times \mathbf{E} - \Omega^2 \mathbf{E} = \mathbf{f}$, which is given in the weak formulation as $\left(\nabla \times \mathbf{E}, \nabla \times \mathbf{E'}\right) - \left( \Omega^2 \mathbf{E}, \mathbf{E'} \right) = \left(\mathbf{f},\mathbf{ E'} \right)$ with a testing function $\mathbf{E'}$. Here it is assumed a given tangential field as boundary condition for $\mathbf{E}$. More specifically, $\mathbf{f}(\mathbf{x})=(\kappa^2 - \Omega^2)(\sin(\kappa x_2), \sin(\kappa x_3), \sin(\kappa x_1))$ on the domain boundaries. 
	For large wavenumber $\Omega$, the problem is highly indefinite and hard to precondition, so typically a direct solver is used.
	We discretize the weak form with first order N\'{e}d\'{e}lec elements using MFEM~\cite{anderson2019mfem}. 
	We use a uniform tetrahedral finite element mesh on a unit cube, resulting in a linear system of size $14,\!827,\!904$ and approximately 24 points per wavelength. The results for $\Omega = 32$ and $\kappa = \Omega / 1.05$ are shown in \cref{fig:Maxwell}.

	\subsection{3D Poisson\label{ssec::P3}}
\begin{figure}[h!]
		\centering
		\begin{subfigure}[t]{.32\textwidth}
			\centering
			\includegraphics[width=\linewidth]{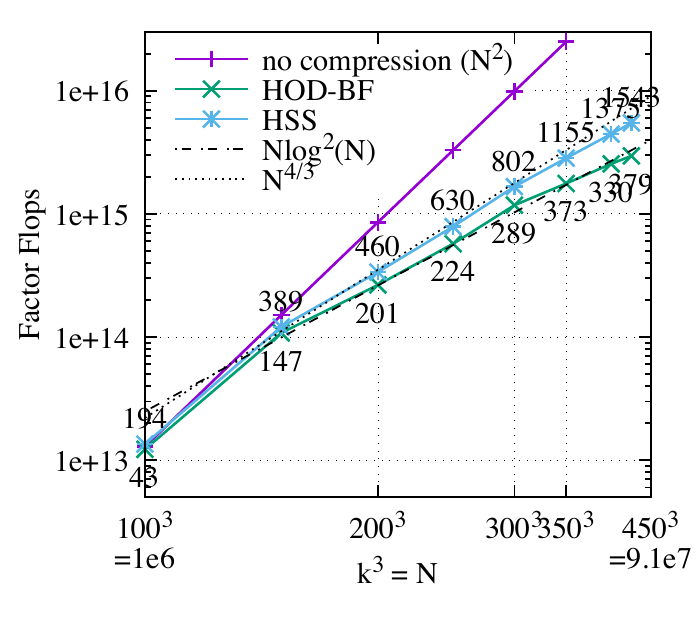}
			\vspace{-50pt}
			\caption{\label{fig:P3_factor_flops}}
		\end{subfigure}
		\begin{subfigure}[t]{.32\textwidth}
			\centering
			\includegraphics[width=\linewidth]{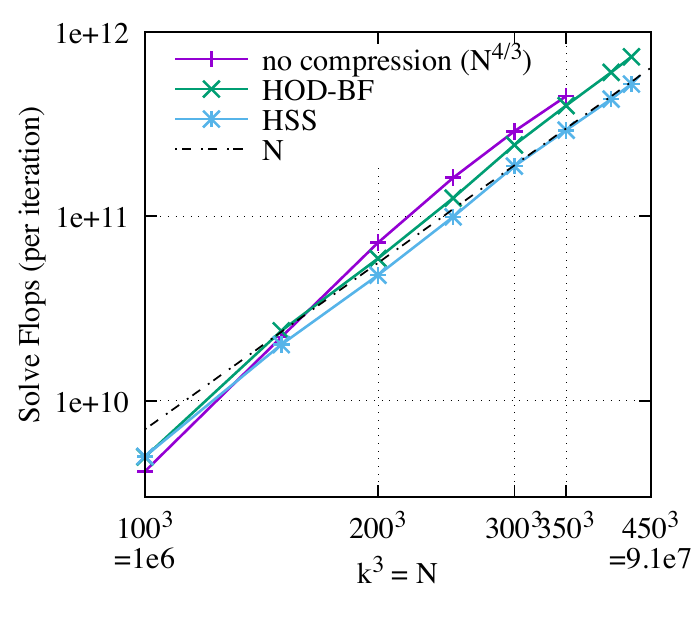}
			\vspace{-50pt}			
			\caption{\label{fig:P3_solve_flops}}
		\end{subfigure}
		\begin{subfigure}[t]{.32\textwidth}
			\centering
			\includegraphics[width=\linewidth]{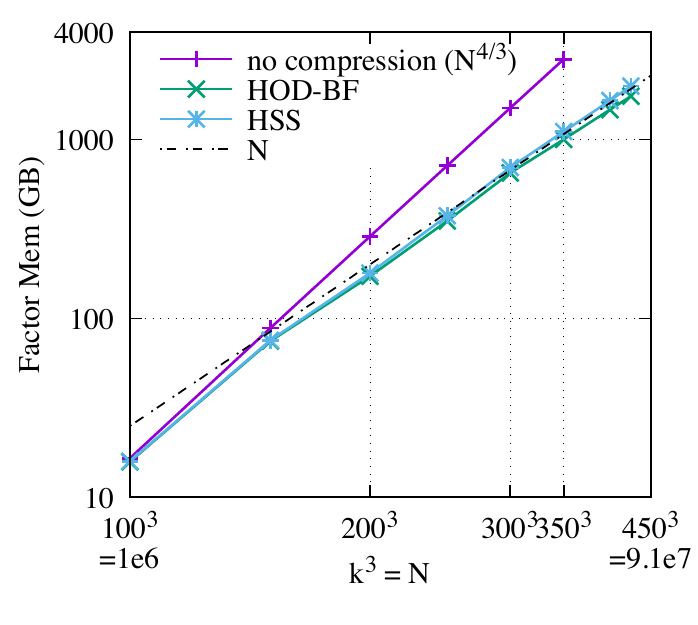}
			\vspace{-50pt}			
			\caption{\label{fig:P3_memory}}
		\end{subfigure}
		\begin{subfigure}[t]{.32\textwidth}
			\centering
			\includegraphics[width=\linewidth]{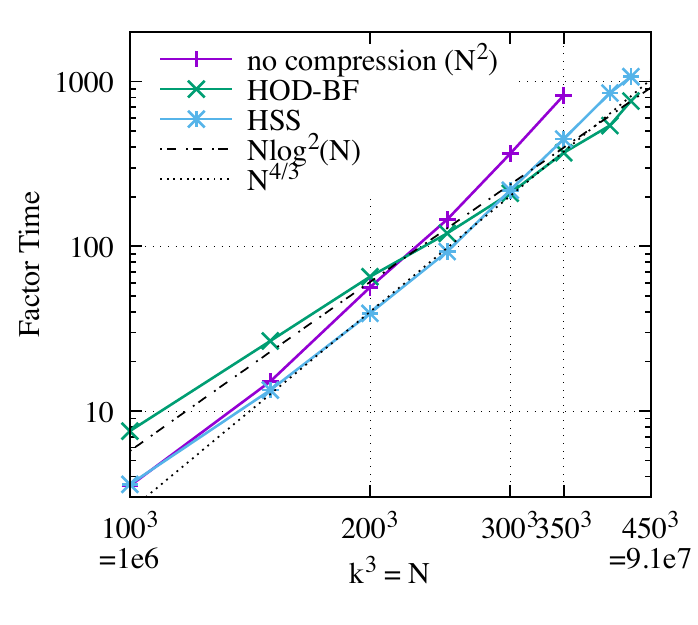}
			\vspace{-50pt}
			\caption{\label{fig:P3_factor_time}}
		\end{subfigure}
		\begin{subfigure}[t]{.32\textwidth}
			\centering
			\includegraphics[width=\linewidth]{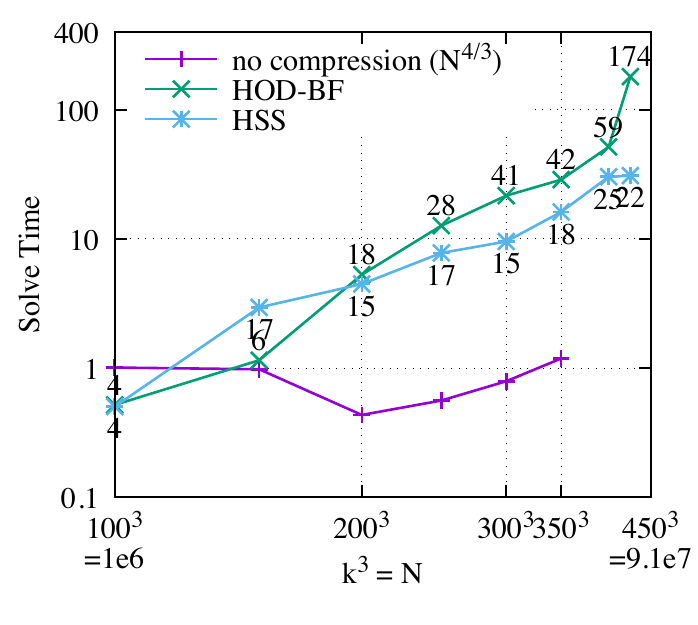}
			\vspace{-50pt}			
			\caption{\label{fig:P3_solve_time}}
		\end{subfigure}
		\begin{subfigure}[t]{.32\textwidth}
			\centering
			\includegraphics[width=\linewidth]{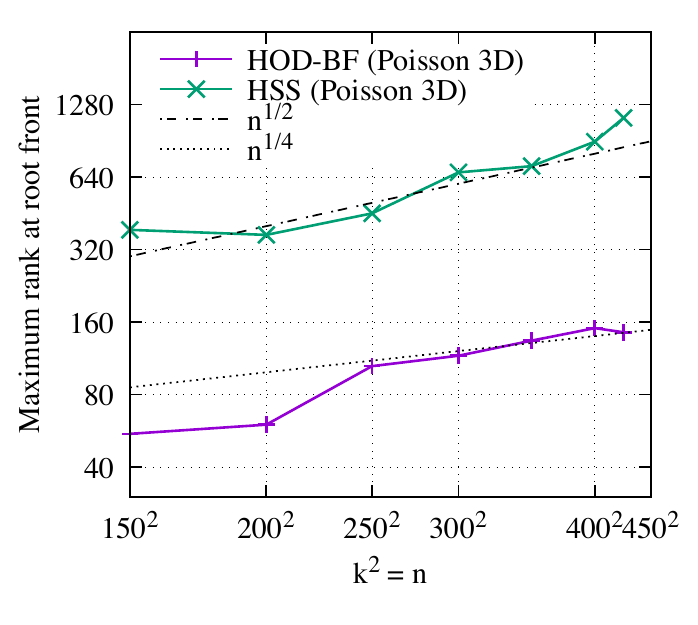}
			\vspace{-50pt}			
			\caption{\label{fig:P3rank}}
		\end{subfigure}	
		\vspace{-10pt}
		\caption{\ylrev{Results for 3D Poisson using the exact solver with iterative refinement, the HOD-BF multifrontal solver with GMRES and compression tolerance $\varepsilon=10^{-3}$, and the HSS multifrontal solver with GMRES and compression tolerance $\varepsilon=10^{-2}$. (a) Flop counts for factorization. The maximum ranks are shown at every datapoint of HOD-BF. (b) Flop counts for solve per iteration in GMRES. (c) Memory usage for the factors (not the peak working memory). (d) CPU time for factorization. (e) CPU time for solve. The number of GMRES iterations are shown at every datapoint of HOD-BF. (f) The maximum ranks encountered in the HSS or HOD-BF representations at the root front.} \label{fig:P3}} 
		\vspace{-20pt}
	\end{figure}

	We solve the Poisson equation on a regular 3D $k^3$ mesh \ylrev{using $64$ compute nodes, with 4 MPI ranks per node and 8 OpenMP threads per MPI process. \cref{fig:P3} shows the factor flop and time, solve flop and time, and factor memory using HOD-BF, HSS and exact multifrontal solvers. The HSS multifrontal solver~\cite{xia2013randomized,ghysels2017robust}, is also implemented in STRUMPACK. The maximum ranks and iteration counts are also shown in \cref{fig:P3_factor_flops} and \cref{fig:P3_solve_time}, respectively. HOD-BF outperforms exact solver and HSS solver respectively for $k>200$ and $k>300$ in terms of factor time, due to quasi-linear complexities predicted by \cref{tab:cc}}. \cref{fig:P3rank} shows that the maximum ranks in the HOD-BF representation, as a function of the size $n$ of the root front, remain much smaller than those in HSS. Note the agreement with~\cref{tab:cc}, which predicts $\mathcal{O}(n^{1/2})$ and $\mathcal{O}(n^{1/4})$ for HSS and HOD-BF respectively. 
	For the \ylrev{$425^3$} problem, the top separator is a \ylrev{$425 \times 425$} plane, corresponding to a \ylrev{$180,\!625^2$} frontal matrix. The largest front is found at the next level, $\ell=1$, and is \ylrev{$270,\!938^2$ ($= 425 \times 425/2 + 425 \times 425$)}. Using HSS, this front is compressed to \ylrev{$2.49 \%$} of the dense storage with a maximum off-diagonal rank of \ylrev{$2856$}, while HOD-BF compresses this front to \ylrev{$0.38 \%$} with a maximum rank of \ylrev{$120$}. 
	

	\section{Conclusion\label{sec:conclude}} 
	This paper presents a fast multifrontal sparse solver for high-frequency wave equations. The solver leverages the butterfly algorithm and its hierarchical matrix extension, HOD-BF, to \ylrev{compress} large frontal matrices. The butterfly representation is computed via fast entry evaluation based on the graph distance, and factorized with randomized matrix-vector multiplication-based algorithms. The resulting solver can attain quasi-linear computation and memory complexity when applied to high-frequency Helmholtz and Maxwell equations. Similar complexities have been analyzed and observed for Poisson equations as well. The code is made publicly available as an effort to integrate the dense solver package ButterflyPACK\footnote{\url{https://github.com/liuyangzhuan/ButterflyPACK}} into the sparse solver package STRUMPACK. \ylrev{To further reduce the overall number of operations and especially the factorization time, a hybrid multifrontal solver which employs HOD-BF for large sized fronts, and BLR or HSS for medium sized fronts is under development.}
	
	\section*{Acknowledgements}
	This research was supported in part by the Exascale Computing Project (17-SC-20-SC), a collaborative effort of the U.S.~Department of Energy Office of Science and the National Nuclear Security Administration, and in part by the U.S.~Department of Energy, Office of Science, Office of Advanced Scientific Computing Research, Scientific Discovery through Advanced Computing (SciDAC) program through the FASTMath Institute under Contract No.~DE-AC02-05CH11231 at Lawrence Berkeley National Laboratory. This research used resources of the National Energy Research Scientific Computing Center (NERSC), a U.S.~Department of Energy Office of Science User Facility operated under Contract No.~DE-AC02-05CH11231.

	\bibliographystyle{plain}
	\bibliography{references}

\end{document}